\documentclass[journal]{IEEEtran}
\usepackage[pdftex]{graphicx}
\DeclareGraphicsExtensions{.pdf,.jpeg,.png,.jpg}
\usepackage{cite}
\usepackage{bm}
\usepackage{float}
\usepackage{array}
\usepackage{wrapfig}
\usepackage{amsmath}
\usepackage{amssymb}
\usepackage{algpseudocode}
\usepackage{color}
\usepackage{cleveref}
\usepackage{graphicx}
\usepackage{subcaption}
\usepackage{mwe}
\newcommand*\diff{\mathop{}\!\mathrm{d}}

\begin{document}
\title{Autonomous Tracking of Intermittent RF Source Using a UAV Swarm}
\author{
	\IEEEauthorblockN{Farshad Koohifar, Ismail Guvenc, and Mihail L. Sichitiu}
	\IEEEauthorblockA{\\Department of Electrical and Computer Engineering, North Carolina State University, Raleigh, NC
    \\Email: \{fkoohif, iguvenc, mlsichit\}@ncsu.edu}
	}
\maketitle
\begin{abstract}
Localization of a radio frequency (RF) transmitter with intermittent transmissions is considered via a group of unmanned aerial vehicles (UAVs) equipped with omnidirectional received signal strength (RSS) sensors. This group embarks on an autonomous patrol to localize and track the target with a specified accuracy, as quickly as possible. The challenge can be decomposed into two stages: 1) estimation of the target position given previous measurements (localization), and 2) planning the future trajectory of the tracking UAVs to get lower expected localization error given current estimation (path planning). For each stage we compare two algorithms in terms of performance and computational load. For the localization stage, we compare a detection based extended Kalman filter (EKF) and a recursive Bayesian estimator. For the path planning stage, we compare steepest descent posterior Cramer-Rao lower bound (CRLB) path planning and a bio-inspired heuristic path planning. Our results show that the steepest descent path planning outperforms the bio-inspired path planning by an order of magnitude, and recursive Bayesian estimator narrowly outperforms detection based EKF.
\end{abstract}

\section{Introduction}
\label{sec:intro}
Enabled by miniaturization of wide variety of sensors and communication modems, unmanned aerial vehicles (UAVs), also colloquially known as drones, are increasingly used for broadband communications, situational awareness, and localization~\cite{6361401 ,7126199,7593027,7124539,7861118,merwaday2016improved}. 
Detecting, localizing, and tracking unauthorized UAVs have been identified as of utmost priority both in military and civilian settings~\cite{6463399,orfanus2016self,hartmann2016uav,guvenc2017detection}. In particular, such UAVs may be used by malicious entities for jamming critical communication links~\cite{6463399} or to collect/hack data from a critical infrastructure, or they may simply be controlled by amateur drone users which may still introduce threats especially if flying near unauthorized areas~\cite{guvenc2017detection}.  
In order to interdict a target UAV, multiple UAVs may need to approach close to the vicinity of the target UAV to launch cyber (e.g. deauthentication~\cite{8088163}) or physical (e.g. using a net to catch the target \cite{7502640,guvenc2018CommMag}) attacks. A variety of sensors on UAVs can be used for localizing and tracking targets, including visual sensors~\cite{7838649}, radio frequency (RF) angle of arrival sensors \cite{7760613}, RF time difference of arrival \cite{7793894}, and received signal strength indicator (RSSI) sensors \cite{5962487}. 

\begin{figure}[t]
	\centering
	\includegraphics[width=0.45\textwidth]{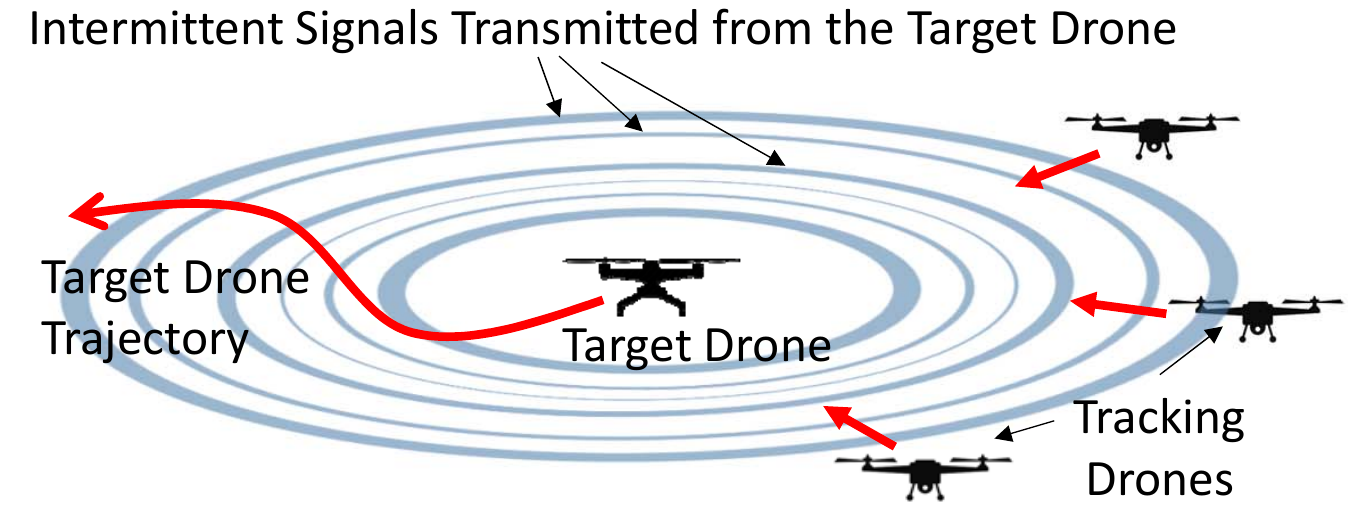}
	\caption{A swarm of UAVs collaborating to localize and track an intermittently transmitting target UAV. The intermittent signals may be jamming signal, video/telemetry signal, or control signal to  a ground station.}
	\label{circ}
\end{figure}

\begin{figure}[t]
	\centering
	\includegraphics[width=0.45\textwidth]{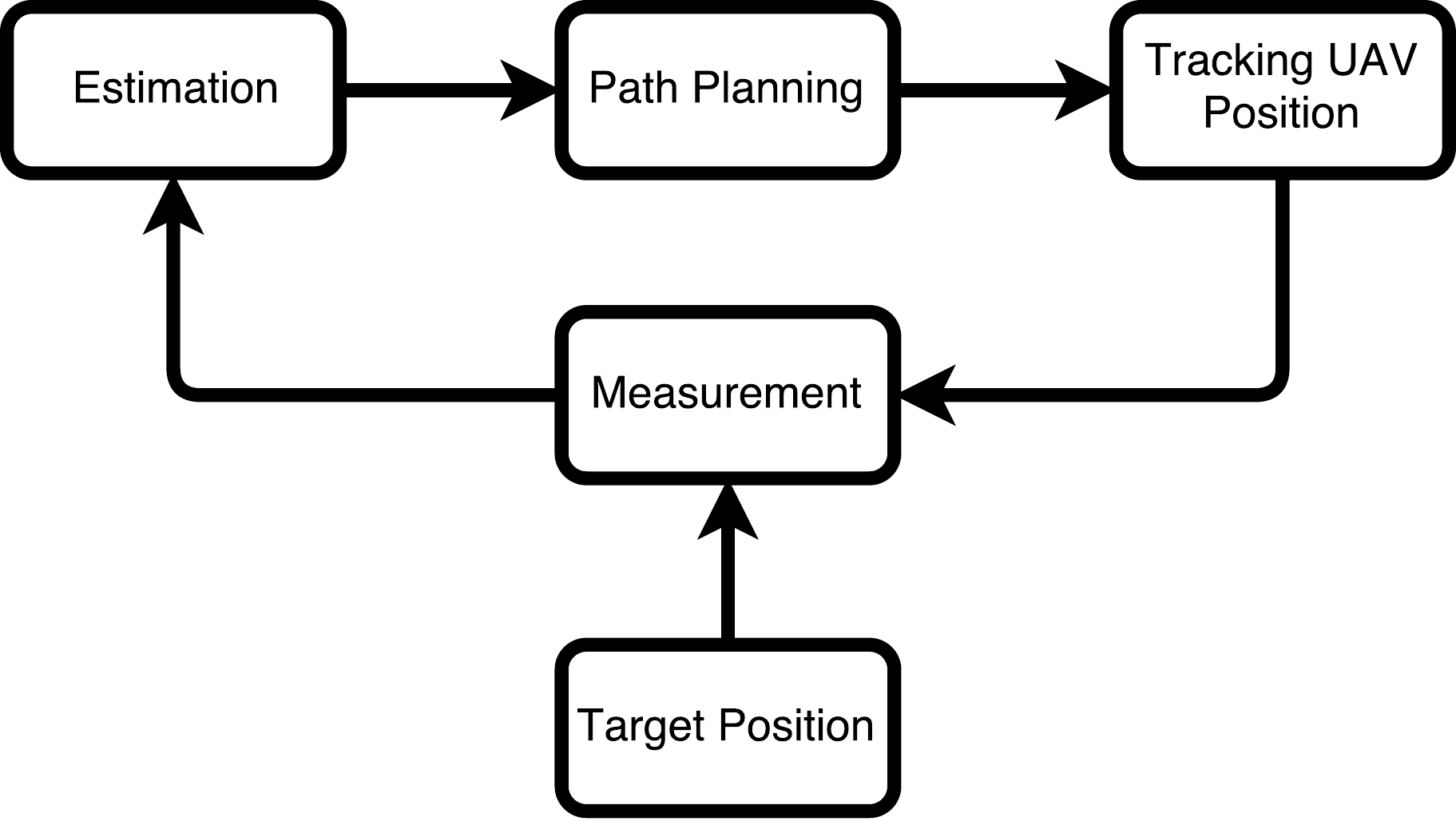}
	\caption{Process diagram for collaborative UAV tracking.}
	\label{overall}
\end{figure}

In our previous work, we have shown that a UAV swarm can accurately localize an RF transmitting target, using inexpensive, omnidirectional RSSI sensors \cite{7555317}. However not all targets transmit continuously, and often, frequency hopping is used in drone communication link to improve reliability and security. Hence, RSSI observations may be only intermittently available, which makes the localization of the target UAV more challenging. In this paper we consider a group of UAVs that are autonomously patrolling a zone of interest in order to find an unauthorized intermittent RF transmitter as illustrated in Fig.~\ref{circ}. The swarm tries to localize the target using an estimation algorithm, and moves the UAVs to get a better measurement in the next step using a path planner algorithm. Fig.~\ref{overall} represents an overall view of the collaborative tracking process, in which each new target location estimate is used to update the path planning.

We compare two different path planning algorithms, namely steepest descent  posterior Cramer-Rao lower bound (CRLB) and a bio-inspired method  described in~\cite{6510083}. In this heuristic bio-inspired approach, the tracking agents move directly towards the last estimate of the target location. The former approach is significantly more computation intensive, while the latter is simply a heuristic based approach with negligible processing burden. Furthermore, we compare two different estimation algorithms, namely a recursive Bayesian estimator and a detection based extended Kalman filter (EKF). Yet again the former is significantly more process intensive, and the latter compromises between optimality and processing power. By comparing these options, we investigate the trade off between performance and computation resources. Relation between measurements, estimation, and path planning steps is summarized in Fig.~\ref{overall}. 

The rest of this paper is organized as follows. In Section~\ref{sec:Literature}, we provide an overview of state of the art. In Section~\ref{sec:system} we describe our approach to model the elements of our system, including stochastic target motion, stochastic target transmission, stochastic channel fading, and measurement noise. At the end of this section, we derive the measurement model that describes the RSSI measurements as an additive unbiased noise to expected value of the measurement. Building upon this model, in Section~\ref{sec:steepest}, we derive the Fisher information matrix (FIM) for the next estimation cycle and derive steepest descent path planning based on the determinant of the FIM. In Section~\ref{sec:EKF}, we describe an EKF that relies on a minimum risk detector to decide if the target was transmitting at this time step, and update the estimation accordingly. In Section~\ref{sec:bayesian}, we describe the Bayesian estimator that does not rely on a separate detector to recursively update the posterior likelihood of target state. In Section~\ref{sec:simulation}, we present our simulation results, and in Section~\ref{sec:conclusion}, we give a summary of our findings.

\section{Literature Review}
\label{sec:Literature}

Several model predictive control based path planning methods are proposed in the literature~\cite{5706736,6965772}. In \cite{7555317}, we proposed a receding horizon path planning method to localize a continuously transmitting RF source. Despite using simple omnidirectional RSSI sensors, we show that path planning based on optimizing D-optimality criterion, paired with an extended Kalman filter as estimator, is a promising approach in autonomous cooperative localization. In \cite{6965772}, localization of an unauthorized transmitter is performed under the constraints of adhering to no-fly zones; in \cite{6965772} UAVs are equipped with electronic support sensors that can detect the angle of arrival of interference, hierarchical model predictive control is used, where the coarse preliminary paths are computed centrally and passed down to all UAVs, to be further optimized locally and individually by each UAV, using a receding horizon optimization technique.

In \cite{7277319}, localization of an RF transmitter using fixed anchor RSSI sensors is proposed. The proposed approach consists of two steps. First, the range between sensors and the target is estimated. Second, localization is achieved using estimated ranges. The node with the highest RSSI is considered as the reference anchor node. The RSSI measurements from all nodes except nodes close to this reference anchor are ignored. Localization is done heuristically, using least square error optimization of range error. 

In \cite{4967999}, mobile targets with intermittent transmission are localized using a state machine to switch between the following four states: 1) global search, 2) approaching target, 3) locating target, and 4) target reacquisition. The cooperation is implemented in a decentralized fashion via 1) a cost function that takes into account the distance that UAV needs to travel to arrive at a location in which it can help the localization, 2) the number of currently helping UAVs at that location, and 3) the number of neighboring UAVs. If the cost function becomes negative, the UAV goes to help; otherwise it will continue its current task. Since the objective is to localize a device with intermittent transmission, the authors have introduced a path planning algorithm that revisits each location regularly while maintaining low energy consumption flight paths.

In \cite{4476759}, the authors investigate localizing simultaneous multiple targets, where time difference of arrival based localization is used by the UAV swarm to estimate the position of the targets. In \cite{Hernandez04optimalsensor}, posterior CRLB is used as the metric to find the optimal trajectories for a group of UAVs using only bearing sensors. Using a computationally efficient method of updating the FIM introduced in \cite{668800}, the dynamic of FIM is decomposed to a nonlinear autonomous response and measurement contribution. The autonomous response entirely depends on the dynamic model of target, and latest FIM, while the measurement contribution depends on the sensor model and position. Through simulations, one step and two step sensor paths are compared, while the number of UAVs varies between one and three. The paper concludes that multi-step multi-UAV scenario results in the best root mean square of error in terms of tracking accuracy.

\section{System Model}
\label{sec:system}
\begin{figure*}
	\begin{center}
	\includegraphics[width=0.8\textwidth]{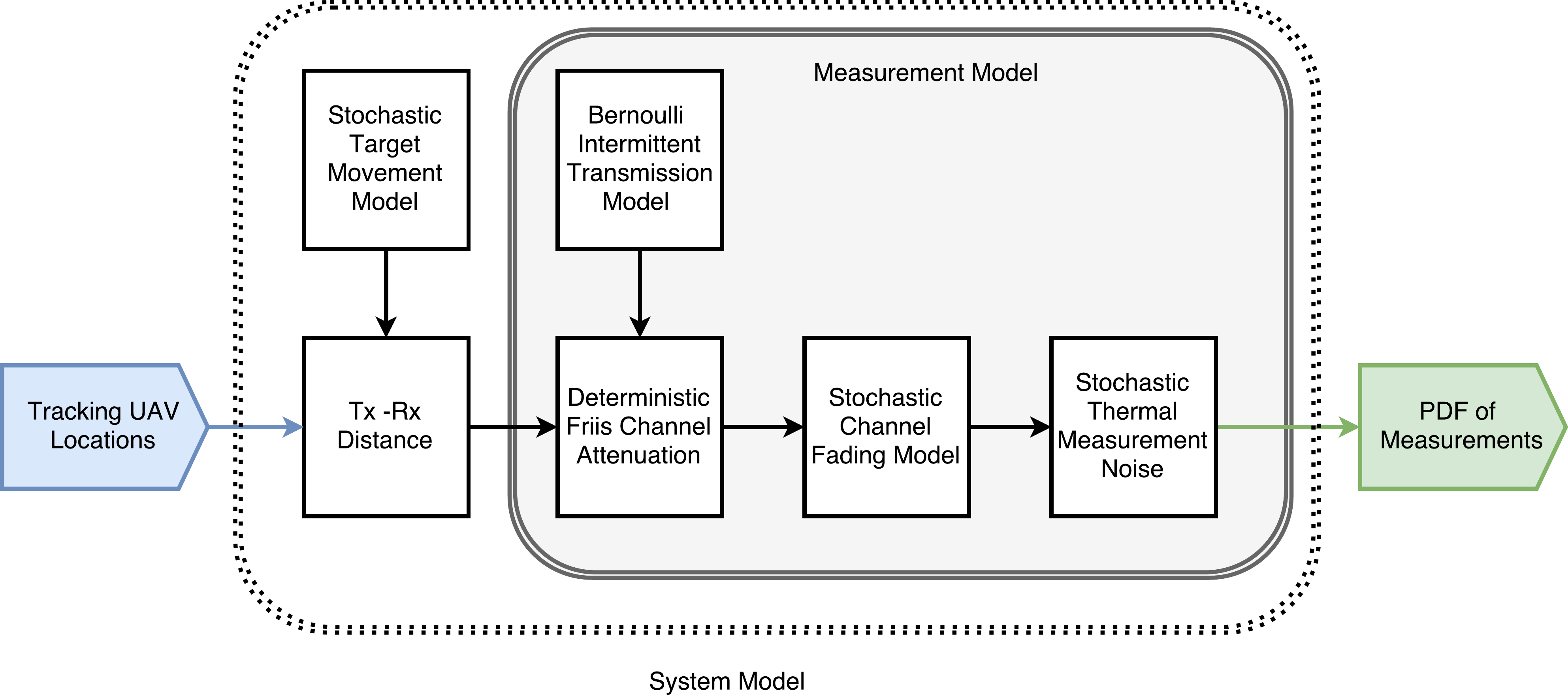}
	\caption{Flowchart summarizing the interaction of stochastic and deterministic subsystem models.}
	\label{fig:systemModel}
\end{center}
\end{figure*}
In this section we describe the models that we are using to represent a group of UAVs equipped with omnidirectional RSSI sensors that are receiving intermittent, omnidirectional transmissions from a stochastically moving target. This system can be modeled by considering each subsystem and their interactions with each other. In Section~\ref{subsec:targetMovement}, we model a moving target with a linear time invariant stochastic process. In Section~\ref{subsec:rssiMeasurement}, we derive the stochastic model for RSSI measurements. Starting from modeling the transmission process of the target with a Bernoulli process, we model the received power at each RSSI sensor after Friis channel attenuation and log-normal fading, given the distance from the target UAV. We then add thermal noise to produce the final measurement model as depicted in Fig.~\ref{fig:systemModel}.

\subsection{Target Movement Model}
\label{subsec:targetMovement}
We use a general stochastic linear time invariant model to describe target movement; our approach works for any stochastic movement model that can be described as a linear model. For the rest of this paper we use an integral Brownian motion model:
\begin{align}
\label{process1}
\mathbf{x}^{(k+1)}&=\mathbf{A}\mathbf{x}^{(k)}+\mathbf{w}^{(k)},\\
\begin{bmatrix}
{x^*_0}^{(k)} \quad  {y^*_0}^{(k)}
\end{bmatrix}^T&=\mathbf{C}\mathbf{x}^{(k)},
\label{process}
\end{align}
where $\mathbf{x}^{(k)} \in \mathbb{R}^M$ is the state vector of the linear movement process and $M$ is the number of target state variables\footnote{In case of the two dimensional dual integral movement model, $M=4$, and the elements of the state vector $\mathbf{x}^{(k)}$ are position and speed in the two horizontal axises.}, $k \in \mathbb{N}$ super script represents the time step, $\mathbf{w}_k  \sim \mathcal{N} (\mathbf{0},\mathbf{Q})$ is the Gaussian process noise with covariance matrix $\mathbf{Q}=\mathbf{E}\{\mathbf{w}^{(k)}{\mathbf{w}^{(k)}}^T\}$, matrix $\mathbf{A}$ is the state transition matrix, and $\mathbf{C}$ is the output matrix that extracts the position of moving target in horizontal and vertical axes, ${x^*_0}^{(k)}$ and ${y^*_0}^{(k)}$. In the double integral model, we have:
\begin{align}
\mathbf{A}=  \begin{bmatrix}
1 \quad 0 \quad \Delta t \quad 0 \\ 0 \quad 1 \quad 0 \quad \Delta t \\ 0 \quad 0 \quad 1 \quad 0 \\ 0 \quad 0 \quad 0 \quad 1 
\end{bmatrix}, \quad \mathbf{C}=  \begin{bmatrix}
1 \quad 0 \quad 0 \quad 0 \\ 0 \quad 1 \quad 0 \quad 0 
\end{bmatrix},
\label{eq:C}
\end{align}
where $\Delta t$ represents time step length.
\subsection{RSSI Measurement Model}
\label{subsec:rssiMeasurement}
In this section we model the observation vector (RSSI measurements) as a deterministic function of unknown parameters (distances from UAVs to target) plus an unbiased measurement noise:
\begin{align}
\mathbf{p}_\text r^{(k)}= \mathbf h^{(k)}+\mathbf{v}^{(k)},
\label{eq:channelModel}
\end{align}
where $\mathbf{p}_\text r^{(k)}\in \mathbb{R}^N$ represents the stochastic measurement vector process resulting from all $N$ UAV measurements at time step $k$, the vector $ \mathbf h^{(k)} \in \mathbb{R}^N$ is the deterministic expected value of the measurement given the unknown parameter vector (distance vector between $N$ UAVs and the target), and the stochastic properties of the measurement are condensed to an unbiased measurement vector noise $\mathbf{v}^{(k)}\in \mathbb{R}^N$. Therefore:           
\begin{align}
\mathbf h^{(k)}&=\mathbf{E}\{\mathbf{p}_\text r^{(k)} \},\\
\mathbf{v}^{(k)}&= \mathbf h^{(k)}-\mathbf{p}_\text r^{(k)}.
\end{align}
To model the measurement process and find $\mathbf h^{(k)}$ and $\mathbf{v}^{(k)}$, we start from modeling a single RSSI sensor that is measuring received power from a single intermittent transmitter, through Friis attenuation, channel fading, and thermal noise. We start from the transmitter and follow the power as it reaches the receiver as depicted in Fig.~\ref{fig:systemModel}. Having the model for a single receiver, we put all $N$ measurements in a vector form, modeling the measurement vector that contains all the measurements in each time step from all UAVs.

We model the intermittent transmission of the target as a Bernoulli process: 
\begin{align}
p_\text t ^{(k)}=s^{(k)}p_\text {on},
\label{bern}
\end{align}
where $p_\text t ^{(k)} \in \mathbb{R}$ represents the intermittent transmitted power process\footnote{Even when the target transmits continuously, due to often used frequency hopping, receivers may observe RSSI only intermittently, due to monitoring only certain portion of the spectrum at a given time.}, $p_\text {on}$ represents transmission power when the transmitter is transmitting, and $s^{(k)}$ is an independent identically distributed Bernoulli process with the probability mass function:
\begin{equation}
\mathcal{L}(s^{(k)}=0) =q, \quad \mathcal{L}(s^{(k)}=1) =1-q,
\end{equation}
where $q$ is the the probability of target not transmitting at any given time step. Each tracking UAV has an omnidirectional RSSI sensor on board. To model the received signal power, we use Friis path loss equation. In the absence of measurement noise, the received power at an arbitrary receiver at distance $d$ away from the transmitter is given by:
\begin{equation}
	p_\textrm{F}^{(k)} =p_\textrm t^{(k)} G_\textrm t G_\textrm r ( {\lambda}/{4 \pi d} )^2,
\end{equation}
where $p_\textrm{F}^{(k)}$ represents the ideal received signal power in absence of shadowing and measurement noise, $G_\textrm t$ is transmitter's antenna gain, $G_\textrm r$ is receiver's antenna gain, and $\lambda$ is the wave length of the transmitted signal. We assume transmitter's power, transmitter's frequency, transmitter's antenna gain, and receiver's antenna gain are known constants. For ease of representation, we define $G$ to represent the constant and known power gain:
\begin{equation}
G \triangleq G_\textrm t G_\textrm r  ( {\lambda}/{4 \pi}  )^2.
\end{equation}
Then in the presence of shadowing effects and without measurement noise, the received signal ($p_\textrm r^* $) is:
\begin{equation}
{p_\textrm r^*}^{(k)} = \frac{p_\textrm t^{(k)} G}{d^2} v_\textrm {sh}^{(k)},
\end{equation}
where $v_\textrm{sh}^{(k)}\sim \textrm{ln}\,\mathcal{N}(0,\,\sigma^{2}_\textrm{sh})$ represents the log normal shadowing effect of the channel. Introduction of received power measurement noise leads to the following:
\begin{equation}
p_\textrm r^{(k)} =p_\textrm t^{(k)} G \frac{1}{d^2} v_\textrm {sh}^{(k)}+p_\textrm {th}^{(k)},
\label{measurementEq}
\end{equation}
where $p_r^{(k)}$ is the measured received power by a single RSSI sensor at time interval $k$, and $p_\textrm {th}^{(k)} \sim \mathcal{N}(\bar p_\textrm {th},\,\sigma^{2}_{p_\textrm {th}})$ is the power measurement noise.

\begin{figure}
	\centering
	\includegraphics[width=0.45\textwidth]{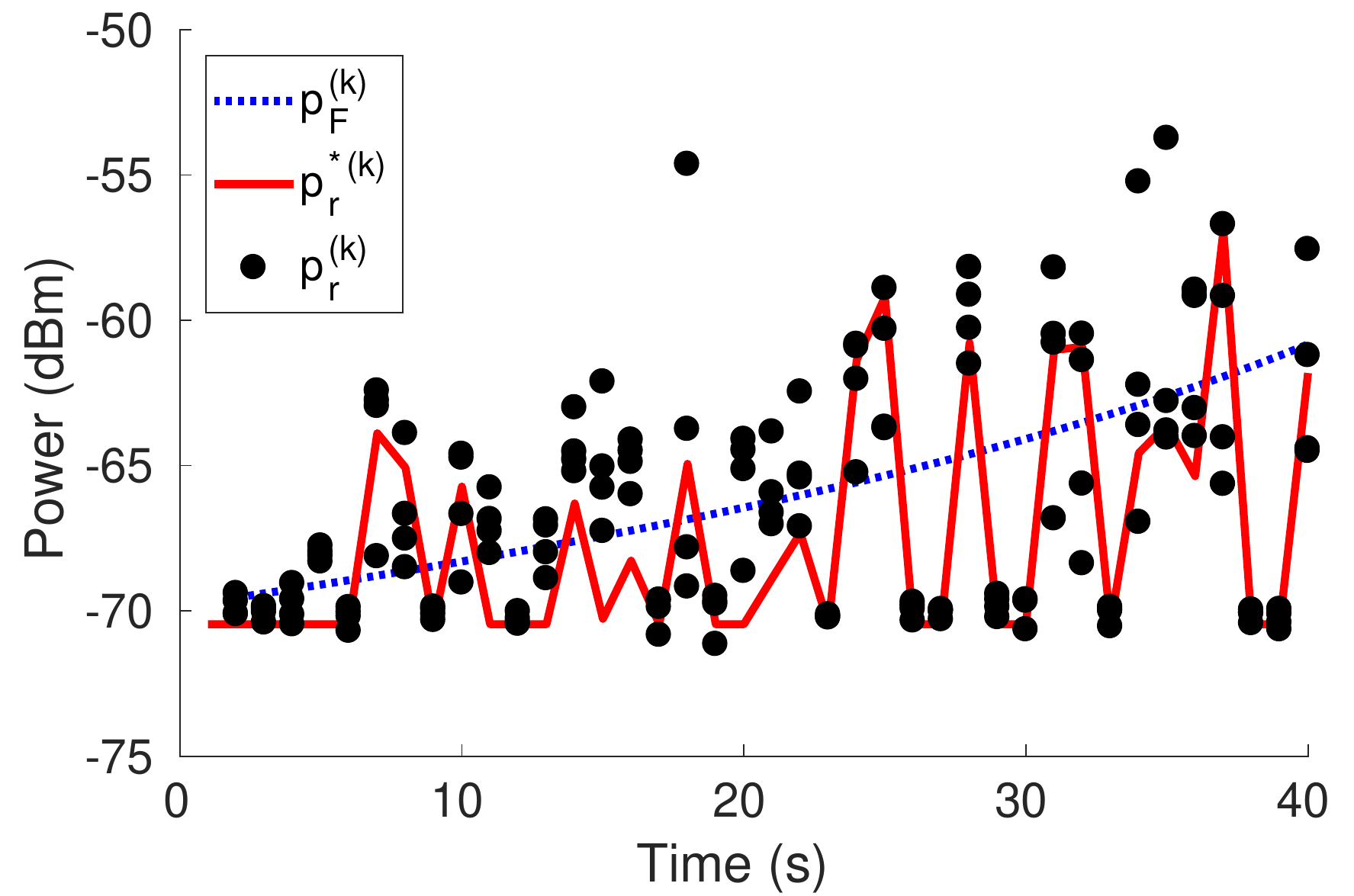}
	\caption{A simulated instance of the RSSI measurement stochastic process. The blue dotted line represents the RSSI in absence of fading and measurement noise. Red line represents an instance of the RSSI with fading but without measurement noise, and dots represent instances of RSSI with fading and measurement noise.}
	\label{0}
\end{figure}

A simulated instance of the measurement model in (\ref{measurementEq}) is presented in Fig.~\ref{0}, where we have simulated 4 UAVs closing in on a stationary target. They start from a distance of 300 meters and move towards the target with a constant speed of 5 m/s. The dashed line represents the received power when target transmits continuously ($q=0$), while there is no shadowing and measurement noise. The solid line shows the effect of turning on the shadowing ($\sigma^{2}_\textrm{sh}=1$) and making the transmission intermittent ($q=0.5$). The black dots are the measurements of the 4 UAV while considering the measurement noise ($\bar p_\textrm {th}=-70\,\textrm{dBM}$, $\sigma_{p_\textrm {th}}=-80\,\textrm{dBm}$).

The measurement model comprises a deterministic component and an unbiased stochastic component. Noting that $s^{(k)}$ and ${v}_\textrm{sh} $ are independent:
\begin{align}
h^{(k)} \triangleq E\{p_\textrm r^{(k)}\}=G\frac{1}{d^2}p_\text {on}{E}\{ s^{(k)}\}  {E}\{v_\textrm{sh}^{(k)} \}+\bar{p}_\textrm{th},
\end{align}
and substituting for expected value of the Bernoulli random variable and log normal random variable, we have:
\begin{align}
h^{(k)}=G\frac{1}{d^2}p_\text {on}(1-q)  e^{\sigma_{\textrm{sh}}^2/2}+\bar{p}_\textrm{th}.
\label{h}
\end{align}
Therefore, the unbiased measurement noise ($v^{(k)} $) is:
\begin{align}
v^{(k)} =\frac{G}{d^2}p_\text {on}[s^{(k)}v_\textrm {sh}^{(k)}+(q-1)  e^{\sigma_{\textrm{sh}}^2/2}]+{p}_\textrm{th}
^{(k)}-\bar {p}_\textrm{th}.
\label{veq}
\end{align}

Equations (\ref{h}) and (\ref{veq}) model a single RSSI sensor. The model captures the expected value of the measurement and the PDF of the measurement noise. In the next step, we consider several RSSI sensors measuring simultaneously on board different UAVs. We consider that the underlying Bernoulli process that governs the transmission is identical for all sensors, but channel fading and thermal noise are independent for each sensor. We stack the resulting measurements in vector form by first defining the distance vector $\mathbf{d}^{(k)} \in \mathbb{R}^N$ between tracking UAVs and the target, where elements of $\mathbf{d}^{(k)}$ are the Euclidean distances between corresponding tracking UAV and the target in time step $k$:
\begin{align}
d_i^{(k)}=\left \lVert
	\mathbf{C}\mathbf{x}^{(k)}-\begin{bmatrix}
	{x_{i}^*}^{(k)} \\ {y_i^*}^{(k)}
	\end{bmatrix}
	\right \rVert.
    \label{distance}
\end{align}
The expected value of the measurements vector ($\mathbf{h}^{(k)}$) is the expected value of each RSSI readout that was derived in (\ref{h}), arranged in vector format corresponding to all $N$ sensors on board of our swarm, which can be written using (\ref{distance}) as:
\begin{equation}
\mathbf h^{(k)}=Gp_\text {on}(1-q)  e^{\sigma_{\textrm{sh}}^2/2}  \begin{bmatrix}
{d_1^{(k)}}^{-2},\quad \ldots, \quad {d_N^{(k)}}^{-2}
\end{bmatrix}^T+\bar{p}_\textrm{th}.
\label{boldh}
\end{equation}
The unbiased noise associated with each sensor's measurement ($\mathbf v^{(k)}$) can be obtained from the unbiased noise derived in (\ref{veq}), for an individual receiver, and can be arranged in vector format corresponding to all $N$ sensors on board of the swarm:
\begin{equation}
\mathbf v^{(k)}=\begin{bmatrix}
Gp_\text {on}[s^{(k)}  {v_\textrm {sh}^{(k)}}_1+(q-1)  e^{\sigma_{\textrm{sh}}^2/2} ] {d_1^{(k)}}^{-2} +{{{p^{(k)}_\textrm{th}}}_1}
\\ \vdots \\
Gp_\text {on}[s^{(k)} { v_\textrm {sh}^{(k)}}_N +(q-1)  e^{\sigma_{\textrm{sh}}^2/2} ]{d_N^{(k)}}^{-2}+{{{{p}^{(k)}_\textrm{th}}}_N}
\end{bmatrix}-\bar{p}_\textrm{th}.
\label{boldv}
\end{equation}
Therefore, the measurement vector ($\mathbf{z}^{(k)}$) is now fully derived based on (\ref{boldh}) and (\ref{boldv}) as follows:
\begin{align}
\mathbf{z}^{(k)} = \mathbf h^{(k)}+\mathbf{v}^{(k)}.
\label{eq:measurement}
\end{align}
This is the measurement model that links UAVs' measurements in (\ref{eq:measurement}) to the target's position through (\ref{distance}), where $\mathbf{z} $ is the measurement vector. Each element of $\mathbf{z} $ is the RSSI measurement of the corresponding tracking UAV.       
\section{Steepest descent path planning algorithm.}
\label{sec:steepest}
In Section~\ref{sec:system}, we have modeled the RSSI measurements from an intermittently transmitting target UAV, observed at tracking UAVs. The stochastic nature of the model dictates how accurate the estimations can be even with optimal estimators. Noisy measurements with low joint likelihood curvature with the parameters of interest (target location) cannot provide additional information to the estimators to allow them to reduce the expected estimation error.  Naturally, the swarm should move in a way that the measurements that they get have good stochastic characteristics and high information. Using the measurement model in (\ref{boldh}) and (\ref{boldv}), swarm paths can be ranked with respect to how much information each path's measurements contain. The maximization of novel information content of measurements is the one and only goal in the steepest descent path planning algorithm.

We make a distinction between information content of a measurement vector and the total posterior information. This key distinction is the motivation for introducing the concept of novel information. We give an example to further clarify this distinction. Suppose the location of a stationary target in x-axis is perfectly known in time step $k$, and we have no information about its location in y-axis. This is our prior information. Any measurement that gives any information about the location of the target in x-axis, is simply providing information that is redundant, because we have perfect information about it already in the prior. In contrast, any information that we gather about target's location in y-axis will increase the total amount of information that we have about the target's location; The total information that we are going to have about target's location after consideration of the new measurements, is posterior information. The difference between prior and posterior information is the novel information content of the measurements. Compare this to the case if we had perfect information in the prior about the location in y-axis, and no information about the location in x-axis. It is clear that the information content of a measurement does not depend on \textit{prior} information, but \textit{novel} information of a measurement directly depends on \textit{prior} information. It is also clear that we are interested in increasing the posterior information, and therefore we are only interested in novel information content of measurements.

Intuitively, a good path planning method should guide the UAVs in such a way that the novel information gathered from next set of measurements is maximized. In other words, given the prior information that we have from the past measurements, find the optimal path that leads to highest amount of posterior information. With this goal in mind, we do not intend to find the measurement set that maximizes the total information in the next measurement, since redundant information is included in that. We intend to find the measurement set that, given our prior information, leads to maximum posterior information.

In this section, we integrate the above intuition in a mathematical framework. In Section~\ref{informationMetric}, we use a well defined metric on information and apply it to our system. In Section~\ref{ppmInformation}, we discuss how prior information and information in a new measurement set combine to form posterior information. Then we discuss how to measure novel information in a measurement set, and optimize for it. We discuss how the posterior information formed in the last time step will loose some of its value and become outdated by the time the next measurement set is available. We discuss how this discounted posterior information will be used as prior for next time step.

\subsection{Information Metric}
\label{informationMetric}
We use FIM as representation of information, which is widely considered to be a good indicator of available information about a stochastic process \cite{784059}, since FIM dictates the maximum achievable estimation accuracy through the CRLB. Given this choice of information metric, we derive a procedure to calculate the amount of posterior information at each time step. To express it formally, we calculate the posterior information about the target state at time step $k$, i.e. the posterior information about $\mathbf x^{(k)}$, given the prior information that we had in time step zero, and all the measurements from time step zero until time step $k$. To be able to mathematically refer to all the measurements that we have from time step zero to time step $k$, we stack all of the measurement vectors up until time step $k$ into a single accumulative measurement vector $\mathbf{z}_\textrm{acc}^{(k)}$:
\begin{align}
\mathbf z^{(k)}_\textrm{acc} \triangleq \begin{bmatrix}
\mathbf z^{(0)}, \quad \ldots ,\quad \mathbf z^{(k)}
\end{bmatrix}^T \in \mathbb R^{kN+N}.
\end{align}
Similarly, we define accumulative target state vector $\mathbf x_\textrm{acc}^{(k)}$, which contains all the states of the target from time step zero to time step $k$:
\begin{align}
\mathbf x_\textrm{acc}^{(k)} \triangleq \begin{bmatrix}
\mathbf x^{(0)}, \quad  \ldots ,\mathbf x^{(k)}
\end{bmatrix}^T \in \mathbb R^{kM+M}.
\end{align}
Given the accumulative measurement vector $\mathbf{z}_\textrm{acc}^{(k)}$, the CRLB establishes the lowest achievable estimation error of the accumulative target state vector $\mathbf{x}_\textrm{acc}^{(k)}$, given known joint probability function $\mathcal{L}(\mathbf z_\textrm{acc}^{(k)},\mathbf{x}_\textrm{acc}^{(k)})$. For any unbiased estimator ${{}\mathbf{\hat x}_\textrm{acc}^{(k)}} =g(\mathbf{z}_\textrm{acc}^{(k)})$, the expected covariance matrix of the estimation error is not better than the CRLB \cite{Kay}:
\begin{align}
	\mathbf{E}\Big \{||g(\mathbf z_\textrm{acc}^{(k)})-\mathbf x_\textrm{acc}^{(k)}||_2^2 \Big \} \geq {\mathbf J_\textrm{acc}^{(k)}}^{-1},
    \label{crlb}
\end{align}
where ${\mathbf{J}_\textrm{acc}^{(k)}}^{-1} \in \mathbb{R}^{(kM+M)\times(kM+M)} $ is the accumulative FIM with the elements given by:
\begin{align}
	{J_{\textrm{acc}_{i,j}}^{(k)} }=-\mathbf{E}\bigg \{\frac{\partial^2 \textrm{ln}( \mathcal{L}(\mathbf{z}_\textrm{acc}^{(k)},\mathbf x_\textrm{acc}^{(k)}))}{\partial {x_\textrm{acc}^{(k)}}_i \partial {x_\textrm{acc}^{(k)}}_j }\bigg \}.
\end{align}
Moreover, ${\mathbf J_\textrm{acc}^{(k)}}^{-1}$ in (\ref{crlb}) represents the posterior information about all states in all steps. As the system progresses through time, the number of unknown parameters in $\mathbf{x}_\textrm{acc}^{(k)}$ and the number of available measurements in $\mathbf{z}_\textrm{acc}^{(k)}$ increase. In next subsection, we use a recursive rule to find the posterior Fisher information of current states of the target, given accumulative measurements.

\subsection{Prior, Posterior, and Measurement Information}
\label{ppmInformation}
Intuitively, at each time step, we want to find the next set point that will lead to the optimal \emph{posterior} CLRB, rather than the entire CRLB. That is to say, we do not value better estimation of target states at previous time steps. Equivalently, we want to set the next measurement path such that the next set of measurements lead to the optimal Fisher information submatrix corresponding only to the target state  in next step. This implies that we do not need to handle and optimize the whole CRLB, but we are only interested in the lower-right $M \times M$ block of the CRLB. Calculation of the entire posterior FIM is both computationally expensive and unnecessary. In this case, where we are only interested in the lower right submatrix, we can use the posterior information submatrix recursion proposed in \cite{668800}. As a result, we will have the posterior CRLB matrix, without the need to handle growing matrices. Using proposition 1 in \cite{668800}, we know that the posterior FIM obeys the recursion:
\begin{align}
\label{Eq:J1}
\mathbf J_{k+1|k}&=-\mathbf D_{2}^T\big(\mathbf J_k+\mathbf D_{1}\big)^{-1}\mathbf D_{2}+\mathbf D_{3},\\
\mathbf J_{k+1}&=\mathbf J_{k+1|k}+\mathbf D_{4},
\label{Eq:J2}
\end{align}
where $\mathbf J_k$ is the FIM corresponding to $\mathbf x^{(k)}$ given $\mathbf{z}_\textrm{acc}^{(k)}$, and $\mathbf J_{k+1|k}$ is the information matrix corresponding to $\mathbf x^{(k+1)}$ given $\mathbf{z}_\textrm{acc}^{(k)}$. The coefficient matrices $\mathbf D_{1}$, $\mathbf D_{2}$, $\mathbf D_{3}$, and $\mathbf D_{4}$ are derived in \cite{668800}:
\begin{align}
\label{d1}
\mathbf D_{1} &\triangleq \mathbf E\{-\Delta^{\mathbf x^{(k)}}_{\mathbf x^{(k)}} \log \mathcal{L}(\mathbf x^{(k+1)}|\mathbf x^{(k)})\},\\
\label{d2}
\mathbf D_{2} &\triangleq \mathbf E\{-\Delta^{\mathbf x^{(k+1)}}_{\mathbf x^{(k)}} \log \mathcal{L}(\mathbf x^{(k+1)}|\mathbf x^{(k)})\},\\
\label{d3}
\mathbf D_{3} &\triangleq \mathbf E\{-\Delta^{\mathbf x^{(k+1)}}_{\mathbf x^{(k+1)}} \log \mathcal{L}(\mathbf x^{(k+1)}|\mathbf x^{(k)})\},\\
\label{d4}
\mathbf D_{4} &\triangleq \mathbf E\{-\Delta^{\mathbf x^{(k+1)}}_{\mathbf x^{(k+1)}} \log \mathcal{L}(\mathbf z^{(k+1)}|\mathbf x^{(k+1)})\},
\end{align}
where we use the denominator layout:
\begin{align}
\Delta^{\boldsymbol \theta}_{\boldsymbol \psi}=\nabla_{\boldsymbol \psi} \nabla^T _{\boldsymbol \theta}, \quad \nabla_{\boldsymbol \theta}=\bigg [ \frac{\partial}{\theta_1},\dots , \frac{\partial}{\theta_r} \bigg]^T.
\end{align}The likelihood functions that we need in order to calculate the coefficient matrices are $ \mathcal{L}(\mathbf x^{(k+1)}|\mathbf x^{(k)})$ and  $\mathcal{L}(\mathbf z^{(k+1)}|\mathbf x^{(k+1)})$. We start from the former (which depends on target movement model) and the later (which depends on measurement model) is derived later in (\ref{key})-(\ref{last}). From (\ref{process1}) we derive $\mathcal{L}(\mathbf x^{(k+1)}|\mathbf x^{(k)})$ :
\begin{multline}
\mathcal{L}(\mathbf x^{(k+1)}|\mathbf x^{(k)})=\textrm{det}(2\pi \mathbf{Q})^{(-1/2)}\times\\
e^{-\frac{1}{2}(\mathbf x^{(k+1)}-\mathbf{A}\mathbf x^{(k)})^T\mathbf{Q}^{-1}(\mathbf x^{(k+1)}-\mathbf{A}\mathbf x^{(k)})}.
\end{multline}
Intuitively, (\ref{Eq:J1}) gives us a method to move information through time. It shows how to calculate the contribution of the information about previous state of the target to information about current state of the target. On the other hand (\ref{Eq:J2}) gives us a method to augment prior information with new information. Applying process model in (\ref{process}) into (\ref{d1}), we can write:
\begin{align}
\mathbf D_{1}= \mathbf E\Big \{\Delta^{\mathbf x^{(k)}}_{\mathbf x^{(k)}}\frac{1}{2}(\mathbf x^{(k+1)}-\mathbf A \mathbf x^{(k)})^T \mathbf Q^{-1}(\mathbf x^{(k+1)}-\mathbf A \mathbf x^{(k)})\Big \},
\end{align}
which implies:
\begin{align}
\mathbf D_{1}=\mathbf A^T \mathbf Q^{-1} \mathbf A.
\label{d11}
\end{align}
Similarly, for (\ref{d2}) and (\ref{d3}) we can write:
\begin{align}
\mathbf D_{2}=-\mathbf A^T\mathbf Q^{-1}, \quad \mathbf D_{3}=\mathbf Q^{-1}.
\label{d21}
\end{align}
In contrast to $\mathbf D_{1}$, $\mathbf D_{2}$, and $\mathbf D_{3}$, which only depend on the process model and handle propagation of information through time, the calculation of $\mathbf D_{4}$ requires applying the measurement model. We have $\mathcal{L}(\mathbf z^{(k+1)}|\mathbf x^{(k+1)})$ equal to $\mathcal{L}(\mathbf z^{(k+1)}|\mathbf h^{(k+1)})$, since $\mathbf h^{(k+1)}$ is a deterministic function of $\mathbf x^{(k+1)}$. Then, using (\ref{eq:measurement}), we can write:
\begin{align}
\mathcal{L}(\mathbf z^{(k+1)}|\mathbf x^{(k+1)})&=\mathcal{L}(\mathbf z^{(k+1)}|\mathbf h^{(k+1)})\nonumber\\
&=\mathcal{L}_{\mathbf v^{(k+1)}} (\mathbf z^{(k+1)}-\mathbf h^{(k+1)}).
\label{key}
\end{align}
Intuitively, this means that the probability of observing a measurement given the process state is equal to probability of measurement noise being equal to difference between the measurement and the expected value of measurement. This can be further simplified by considering the on and off possibilities for the Bernoulli process. We define $\alpha$ as the probability of observing a measurement noise when transmitter has transmitted, and $\beta$ when it has not:
\begin{align}
\label{alpha}
	\alpha &\triangleq\mathcal{L}_{\mathbf v^{(k+1)}}(\mathbf z^{(k+1)}-\mathbf h^{(k+1)}|s^{(k+1)}=1),\\
	\beta &\triangleq\mathcal{L}_{\mathbf v^{(k+1)}}(\mathbf z^{(k+1)}-\mathbf h^{(k+1)}|s^{(k+1)}=0),
    \label{beta}
\end{align}
where we have:
\begin{align}
\mathcal{L}(\mathbf z^{(k+1)}|\mathbf h^{(k+1)})&=(1-q)\alpha+q \beta.
\end{align}
We define new dependent vectors $\mathbf{k}$ and $\mathbf{l}$ that describe measurement noise when the transmitter was on and off:
\begin{align}
\mathbf k &\triangleq \mathbf v^{(k+1)}|s^{(k+1)}=1,\\
\mathbf l&\triangleq \mathbf v^{(k+1)}|s^{(k+1)}=0,
\end{align}where from (\ref{boldv}), we can write:
\begin{align}
k_{i}(v_\textrm {sh},{p}_\textrm{th})&=Gp_\text {on}[ v_\textrm {sh}+(q-1)  e^{\sigma_{\textrm{sh}}^2/2} ]d_ i^{-2}+{{p}_\textrm{th}}-\bar{p}_\textrm{th},\\
l_{i}({p}_\textrm{th})&= Gp_\text {on}(q-1)  e^{\sigma_{\textrm{sh}}^2/2} d_ i^{-2}+{{p}_\textrm{th}}-\bar{p}_\textrm{th}.
\end{align}
We assume that the measurement process at each tracking UAV is independent, and as a result, we can rewrite (\ref{alpha}) and (\ref{beta}) as:
\begin{align}
\label{alpha}
	\alpha &= \prod_{i=1}^N \mathcal{L}_{k_{i}}(z_i^{(k+1)}-h_i^{(k+1)}),\\
    \label{beta}
	\beta &= \prod_{i=1}^N \mathcal{L}_{l_{i}}(z_i^{(k+1)}-h_i^{(k+1)})~,
\end{align}
where since the distribution of $v_\textrm{sh}\sim \textrm{ln}\,\mathcal{N}(0,\,\sigma^{2}_\textrm{sh})$ and $p_\textrm {th} \sim \mathcal{N}(\bar p_\textrm {th},\,\sigma^{2}_{p_\textrm {th}})$ are known:
\begin{align}
\zeta   &\triangleq (1-q)  e^{\sigma_{\textrm{sh}}^2/2}+ \frac{z_i({k+1})-h_i({k+1})-{{p}_\textrm{th}}+\bar{p}_\textrm{th}}{Gp_\text {on} d_ i^{-2}}\\
\begin{split}
&\mathcal{L}_{k_{i}}(z_i^{(k+1)}-h_i^{(k+1)})\\
&= \frac{d_ i^{2}}{Gp_\text {on}}  \int_{0}^{\infty}\mathcal{L}_{v_\textrm {sh} }(   \zeta)\mathcal{L}({p}_\textrm{th})\diff p_\textrm{th}\\
&=\frac{d_ i^{2}}{Gp_\text {on}}  \int_{0}^{\infty}       \frac{1}{\zeta \sigma_\textrm{sh} \sqrt[]{2\pi}}e^{-\frac{(\ln \zeta)^2 }{2 \sigma_\textrm{sh}^2}}              
\frac{1}{\sqrt{2\sigma^2_\textrm{th}\pi}}e^{-\frac{({p}_\textrm{th}-\bar{p}_\textrm{th})^2}{2\sigma^2_\textrm{th}}} \diff {p}_\textrm{th},
\end{split}\\
\eta &\triangleq z_i^{(k+1)}-h_i^{({k+1})}- Gp_\text {on}(q-1)  e^{\sigma_{\textrm{sh}}^2/2} d_ i^{-2},\\
\begin{split}
\mathcal{L}_{l_{i}}(z_i^{({k+1})}-h_i^{({k+1})})=\frac{1}{\sqrt{2\sigma^2_\textrm{th}\pi}}e^{-\frac{\eta ^2 }{2\sigma^2_\textrm{th}}}
\end{split}.
\label{last}
\end{align}

At this point we have derived $\mathcal{L}(\mathbf z^{(k+1)}|\mathbf x^{(k+1)})$, but to get $\mathbf D_{4}$ from (\ref{d4}), we need to numerically calculate the expected value of the Hessian of this likelihood function. To do so we use Monte Carlo expectation on numerical Hessian. We pick a random measurement noise instance ($\hat{\mathbf v}^{(k+1)}$) from its distribution, given $\mathbf x^{(k+1)}$, then calculate a corresponding Hessian:
\begin{align}
\hat{\mathbf H}&=\Delta^{\mathbf x^{(k+1)}}_{\mathbf x^{(k+1)}} \log \mathcal{L}(\mathbf z^{(k+1)}|\mathbf x^{(k+1)}) \Big|_{\hat{\mathbf v}^{(k+1)}},
\end{align}
whose individual elements can be calculated as:
\begin{align}
\begin{split}
\hat{h}_{i,j}&\approx \big (\log \mathcal{L}(\mathbf z^{(k+1)}|\mathbf x^{(k+1)}+\delta_i+\delta_j)\\
&-\log \mathcal{L}(\mathbf z^{(k+1)}|\mathbf x^{(k+1)}+\delta_i-\delta_j)\\
&-\log \mathcal{L}(\mathbf z^{(k+1)}|\mathbf x^{(k+1)}-\delta_i+\delta_j)\\
&+\log \mathcal{L}(\mathbf z^{(k+1)}|\mathbf x^{(k+1)}+\delta_i+\delta_j)\big )/4\Delta^2 \Big|_{\hat{\mathbf v}^{(k+1)}}~,
\end{split}
\end{align}
where $\Delta$ is derivative step size, and $\delta_i$ is the $i^\mathrm{th}$ unit vector scaled to $\Delta$. Monte Carlo expectation of this Hessian leads to numerical evaluation of $\mathbf D_{4}$ since:
\begin{align}
\mathbf D_{4}&=-\mathbf{E}\{\hat{\mathbf H}\},
\end{align}
The algorithm is presented in Fig.~\ref{numerical}.
\begin{figure}
	\begin{algorithmic}[1]
		\Require $\hat{\mathbf{x}}_{k-1|k-1},\mathbf{P}_{k-1|k-1},\mathbf z_k$
		\Procedure{SteepestDescentPathPlanning}{}
        \State $\mathbf D_{1}\gets\mathbf A^T \mathbf Q^{-1} \mathbf A$
		\State $\mathbf D_{2}\gets-\mathbf A^T\mathbf Q^{-1}$
		\State$ \mathbf D_{3}\gets\mathbf Q^{-1}$
		\For{$\mathbf{s}$ \textbf{in} $\mathbf{S}$}
			\State $\mathbf {x^*}^{(k+1)}\gets \mathbf {x^*}^{(k)}+\cos (\mathbf{s})\Delta L$
			\State $\mathbf {y^*}^{(k+1)}\gets \mathbf {y^*}^{(k)}+\sin (\mathbf{s})\Delta L$
       	 	\State $\mathbf D_{4}\gets $ \Call{doExpectation}{$.$}
        	\State $\mathbf J(k+1)=-\mathbf D_{2}^T\big(\mathbf J(k)+\mathbf D_{1}\big)^{-1}\mathbf D_{2}+\mathbf D_{3}+\mathbf D_{4}$
        	\State $c(\mathbf{s} )\gets |\mathbf J(k+1)|$
		\EndFor
		\State $\mathbf s_\textrm{opt}=\arg \max c(\mathbf{s} )$
		\EndProcedure	
\end{algorithmic}
\begin{algorithmic}[1]
    \Procedure{doExpectation}{}
    \State $\textrm{H}\gets \mathbf{0}_{r,r}$
     \For{$\textrm{try}$ \textbf{in} $1:t$}
    \State $\hat{\mathbf v}^{(k+1)} \gets$ \Call{errorInstance}{.}
    \State $\hat{\textrm{H}}\gets \hat{\textrm{H}}- \textrm{H}/t$
    \EndFor
\EndProcedure	
\Procedure{$\mathcal{L}(\mathbf z^{(k+1)}|\mathbf x^{(k+1)}) \Big|_{\hat{\mathbf v}^{(k+1)}}$}{}
    		\State $\alpha \gets \beta \gets 1$
		\For{$i$ \textbf{in} $1:N$}
        \State $\zeta  \gets (1-q)  e^{\sigma_{\textrm{sh}}^2/2}+ \frac{z_i({k+1})-h_i({k+1})-{{p}_\textrm{th}}+\bar{p}_\textrm{th}}{Gp_\text {on} d_ i^{-2}}$
		\State $\alpha \gets \alpha \frac{d_ i^{2}}{Gp_\text {on}}  \int_{0}^{\infty}       \frac{1}{\zeta \sigma_\textrm{sh} \sqrt[]{2\pi}}e^{-\frac{(\ln \zeta)^2 }{2 \sigma_\textrm{sh}^2}}              
\frac{1}{2\sigma^2_\textrm{th}\pi}e^{-\frac{({p}_\textrm{th}-\bar{p}_\textrm{th})^2}{2\sigma^2_\textrm{th}}}   \diff p_\textrm{th}$
		\State $\eta \gets z_i^{(k+1)}-h_i^{(k+1)}- Gp_\text {on}(q-1)  e^{\sigma_{\textrm{sh}}^2/2} d_ i^{-2}$
        \State $\beta \gets \beta \frac{1}{2\sigma^2_\textrm{th}\pi}e^{-\frac{\eta ^2}{2\sigma^2_\textrm{th}}} $
		\EndFor
        \State $\mathcal{L}(\mathbf z^{k+1}|\mathbf h^{k+1})\gets (1-q)\alpha+q \beta$
\EndProcedure	
\end{algorithmic}
	\caption{Steepest descent path planning.}\label{numerical}
\end{figure}
Having numerical evaluation of $\mathbf D_{4}$, the computationally efficient recursion of Fisher information submatrix can be used.

\subsection{Path Planning}

Intuitively, we desire a path planning approach that leads to the optimal FIM. Since matrix spaces do not have a unique norm, our intuitive criterion of finding the optimal FIM is ambiguous. One of the most common approaches is to minimize the determinant of posterior CRLB or equivalently, maximize the determinant of the FIM \cite{7468562}. This criterion is called D-optimality criterion. With this criterion in mind, we define $\mathbf s \in [-\pi,\ \pi]^N$ as the vector of UAV directions that is dictated by path planning. The next Fisher information submatrix is a function of our path planning decision on which direction the UAVs should travel for the next time step. To note this dependency explicitly, we use the notation $\mathbf J_{k+1}(\mathbf s)$ for the next Fisher information submatrix and write the UAV path planning problem as:
\begin{align}
	\mathbf s_\textrm{opt}=\arg \max_\mathbf s |\mathbf J_{k+1}(\mathbf s)|,
\end{align}
where $\mathbf s_\textrm{opt}$ is the optimal direction that the UAVs need to take, and $\mathbf J(\mathbf s)$ is the FIM corresponding to states of the target at next time step. To do this optimization, we need to find the gradient of determinant of the FIM. The determinant of FIM in this case is a scalar function over the field of movement directions of UAVs. Note that at any time step, we may have already acquired some measurements and therefore we may have some amount of information about the location of the target and the FIM should take previous measurements into account and only consider the new information that will be generated.

The path planning stage optimizes the information around a point in target state space. Choosing that point in target state space to optimize the information around it, needs a target state estimator to estimate where in the state space our target is (e.g. location and velocity for double integral movement model). In the next two sections we introduce two specific  state estimators that can do this task: detection based EKF and Bayesian estimator.

\section{Detection Based EKF}
\label{sec:EKF}
We introduce detection based EKF as a target location estimation approach. Intuitively, if we knew that the target has transmitted, we know how to update the estimation of the target location, given prior estimation and its expected error covariance. We could use EKF to do the update. Similarly, if we knew target did not transmit, we know how to update our estimation, solely by predicting current target state from its previous estimation. Therefore one way of localization is to first use a detector to detect if the target has transmitted, and decide how to update the estimation based on detector's output. We refer to this estimator as \textit{detection based EKF}.

Detection based EKF relies on a central collective decision to be made based on measurements in current time step to detect if the transmitter has transmitted or was silent in this time period. Derivation of optimal detection method requires a discussion of propagation of error through time, when the EKF does a correct detection, false alarm, misdetection, and correct rejection. Therefore we describe how error propagates in each of these scenarios, then use them to do optimal detection that reduces the overall determinant of estimation error covariance matrix.

\begin{figure}
	\begin{algorithmic}[1]
		\Require $\hat{\mathbf x}^{k-1|k-1},\mathbf P^{k-1|k-1},\mathbf z_k$
		\Procedure{Detection Based EKF}{}
		\State $\hat{\mathbf{x}}^{(k|k-1)} \gets \mathbf{A}\hat{\mathbf x}^{(k-1|k-1)}$
		\State $\mathbf P^{(k|k-1)} \gets \mathbf{A}\mathbf P^{(k-1|k-1)}\mathbf{A}^T+\mathbf Q$
		\State $\tilde{\mathbf{y}}^{(k)} \gets \mathbf z^{(k)}-\mathbf{h}(\hat{\mathbf{x}}^{(k|k-1)})$
		\State $\mathbf{S}^{(k)} \gets {\mathbf{H}^{(k)}}^T\mathbf{P}^{(k|k-1)} \mathbf{H}^{(k)}+\mathbf{R}$
		\State $\mathbf{K}^{(k)} \gets \mathbf{P}^{(k|k-1)} \mathbf{H}^{(k)} {\mathbf{S}^{(k)}}^{-1}$
		\State $U_{11}\gets\begin{vmatrix}(\mathbf{I}-\mathbf{K}^{(k)}{\mathbf{H}^{(k)}}^T)\mathbf{P}^{(k|k-1)} \end{vmatrix},$
		\State
		$U_{12}\gets\begin{vmatrix}\mathbf{P}^{(k|k-1)}+ \mathbf{K}^{(k)} \tilde{\mathbf{y}}^{(k)}(\mathbf{K}^{(k)} {\tilde{\mathbf{y}}^{(k)}})^T\end{vmatrix}$
		\State
		$U_{21}\gets\begin{vmatrix}\mathbf{P}^{(k|k-1)} \end{vmatrix}$
		\State		$U_{22}\gets\begin{vmatrix}\mathbf{P}^{(k|k-1)} \end{vmatrix}$
		\State $\tau^{(k)} \gets \frac{p(z|\mathcal{H}_1)}{p(z|\mathcal{H}_0)}>\frac{(U_{12}-U_{22})q}{(U_{21}-U_{11})(1-q)}$
		\If{$\tau_k$}
		\State $\mathbf{P}^{(k|k)} \gets (\mathbf{I}-\mathbf{K}^{(k)} {\mathbf{H}^{(k)}}^T)\mathbf{P}^{(k|k-1)}$
		\State $\hat{\mathbf{x}}^{k|k} \gets \hat{\mathbf{x}}^{(k|k-1)}+\mathbf{K}^{(k)} \tilde{\mathbf{y}}^{(k)}$
		\Else
        \State $ \mathbf{P}^{(k|k)} \gets {\mathbf{P}}^{(k|k-1)}$
		\State $	 \hat{\mathbf{x}}^{(k|k)} \gets \hat{\mathbf{x}}^{(k|k-1)}$
		\EndIf
		\State \Return $\hat{\mathbf{x}}^{(k|k)},\mathbf{P}^{(k|k)}$
		\EndProcedure	
	\end{algorithmic}
	\caption{Detection based EKF algorithm for target UAV location estimation.}\label{euclid}
\end{figure}

If a transmission is detected, the current measurement vector is used to update the EKF estimation. Otherwise, we update the state estimation using one step ahead prediction and update the state estimation error covariance matrix $\mathbf{P}^{(k|k)}$ accordingly as described in Fig.~\ref{euclid}. We use $\tau^{(k)}$ to represent detection decision and $\hat{\mathbf{x}}^{(k|k-1)}$ is the prediction of target UAV state at time step $k$ given measurements up until time step $k-1$. This prediction's covariance is represented with $\mathbf{P}^{(k|k-1)}$, observed residual is represented with $\tilde{\mathbf{y}}^{(k)}$, residual covariance is represented with $\mathbf{S}^{(k)}$, Kalman gain is represented with $\mathbf{K}^{(k)}$, $\mathbf{H}^{(k)}$ is the Jacobian of $\mathbf{h}^{(k)}$ in \eqref{boldh} which can be written as:
\begin{align}
	\mathbf{H}^{(k)}&=\frac{\partial \mathbf{h}}{\partial \mathbf{x}} \Big|_{\hat{\mathbf x}^{(k|k-1)}}\\
  \begin{split}
  h_{i,j}&=Gp_\text {on}(1-q)  e^{\sigma_{\textrm{sh}}^2/2} d_j^{-4}  \Big(c_{1,i} \big(\mathbf C_1 \mathbf x^{(k)}-{x^*_j}^{(k)}\big)\\
  &+ c_{2,i}\big(\mathbf C_2 \mathbf x^{(k)}-{y^*_j}^{(k)}\big)\Big)\Big|_{\hat{\mathbf{x}}^{(k|k-1)}},
  \end{split}
\end{align}
and $\mathbf{R}$ is the measurement error covariance matrix:
\begin{align}
\mathbf{R}=\mathbf{E}\{\mathbf{v}^{(k)}{\mathbf{v}^{(k)}}^T\},
\end{align}
where the individual elements of the matrix ${\bf R}$ are:
\begin{align}
r_{i,j}=
\begin{cases} 
(Gp_\text {on})^2d_i^{-2} d_j^{-2} q(1-q)e^{\sigma_{\textrm{sh}}^2}  \hfill &  \textrm{,if }i\neq j  \\
\hfill(Gp_\text {on})^2d_i^{-4} (1-q)(e^{\sigma_{sh}^2}-1)e^{\sigma_{sh}^2}\hfill & \\
\hfill +\sigma_{th}^2-(1-q)^{2}e^{\sigma_{\textrm{sh}}^2} &\textrm{,if } i=j \\
\end{cases}.
\end{align}

Having calculated the consequence of each decision on estimation procedure, we turn into finding an optimal decision making method. The two hypotheses we need to choose between correspond to target having transmitted in time step $k$ or not:
\begin{align}
	\mathcal{H}_0: s^{(k)}=0, \quad \mathcal{H}_1: s^{(k)}=1.
\end{align}
Intuitively, the detector should not have a constant false alarm rate or fixed probability of detection. The detector should take into account the expected cost of its actions, and adapt its threshold accordingly to minimize the total expected cost. Using the determinant of estimation error covariance matrix as cost function, we derive the cost associated with each of the possible outcomes of our detection method as follows:
\begin{align}
U_{11}&=\begin{vmatrix}(\mathbf{I}-\mathbf{K}^{(k)}{\mathbf{H}^{(k)}}^T)\mathbf{P}^{(k|k-1)} \end{vmatrix},\\
U_{12}&=\begin{vmatrix}\mathbf{P}^{(k|k-1)}+ \mathbf{K}^{(k)} \tilde{\mathbf{y}}^{(k)}(\mathbf{K}^{(k)} \tilde{\mathbf y}^{(k)})^T\end{vmatrix},\\
U_{21}&=U_{22}=\begin{vmatrix}\mathbf{P}^{(k|k-1)} \end{vmatrix},
\end{align}
where $U_{11}$ is cost of correct detection, $U_{12}$ is cost of false alarm, $U_{21}$ is cost of misdetection, and $U_{22}$ is cost of correct rejection. Having calculated the costs associated with each decision outcome, we use optimal Bayesian cost detector \cite{kayd} to decide $\mathcal{H}_1$ if the following condition is satisfied:
\begin{align}
	\frac{p(z|\mathcal{H}_1)}{p(z|\mathcal{H}_0)}>\frac{(U_{12}-U_{22})q}{(U_{21}-U_{11})(1-q)}.
    \label{LRT}
\end{align}

\section{Bayesian Estimation} \label{sec:bayesian}

As an alternative estimator for target location, we can use a Bayesian estimator. In contrast to detection based EKF estimator in Section~\ref{sec:EKF}, Bayesian estimator does not rely on explicitly detecting if target has transmitted in the last time step or not. In the Bayesian estimation approach, we aim to achieve minimum mean square error:
\begin{align}
\hat{\mathbf x}^{(k|k)}=\textrm{arg} \min_{\tilde{\mathbf{x}}} \mathbf{E}\{(\tilde{\mathbf{x}}-\mathbf{x}^{(k)})^2\},
\end{align}
with the solution:
\begin{align}
\hat{\mathbf{x}}^{(k|k)}&=\mathbf{E}\{\mathbf{x}^{(k)}|\mathbf{z}^{(k)},\ldots,\mathbf{z}^{(0)}\},\\
&=\int \mathbf{x}^{(k)} \mathcal{L} (\mathbf{x}^{(k)}|\mathbf{z}^{(k)},\ldots,\mathbf{z}^{(0)})\diff \mathbf{x}^{(k)}.
\label{eq:mmse}
\end{align}
Therefore, we need to calculate $\mathcal{L}(\mathbf{x}^{(k)}|\mathbf{z}^{(k)},\mathbf{z}^{k-1},\ldots,\mathbf{z}^{(0)})$, which is the likelihood function of the target state conditioned on all acquired measurements. Calculation of this likelihood  function can be done recursively. First, using the conditional likelihood function of the last step ($\mathcal{L} (\mathbf{x}^{(k-1)}|\mathbf{z}^{(k-1)},\ldots,\mathbf{z}^{(0)})$), we predict the conditional likelihood function of the target state:
\begin{align}
&\mathcal{L} (\mathbf{x}^{(k)}|\mathbf{z}^{(k-1)},\ldots,\mathbf{z}^{(0)})=\nonumber\\
&\int \mathcal{L} (\mathbf{x}^{(k)}|\mathbf{x}^{(k-1)})\mathcal{L} (\mathbf{x}^{(k-1)}|\mathbf{z}^{(k-1)},\ldots,\mathbf{z}^{(0)})\diff \mathbf{x}^{(k-1)},
\label{prior}
\end{align}
where from (\ref{process1}), we can write:
\begin{multline}
\mathcal{L} (\mathbf{x}^{(k)}|\mathbf{x}^{(k-1)})=\\
|2\pi \mathbf{Q}|^{-\frac{1}{2}}e^{-\frac{1}{2}(\mathbf{x}^{(k)}-\mathbf{A}\mathbf{x}^{(k-1)})^T\mathbf{Q}^{-1}(\mathbf{x}^{(k)}-\mathbf{A}\mathbf{x}^{(k-1)})}.
\end{multline}

\begin{figure}
	\begin{algorithmic}[1]
    	\Require $\textrm{Posterior}(\mathbf{x}^{(k-1)}),\mathbf z^{(k)}$
		\Procedure{Bayesian Estimator}{}
        \For{\textbf{all} $\mathbf{x}^{(k)}$}
			 \State $\textrm{Prior}(\mathbf{x}^{(k)})\gets \sum\limits_{\mathbf{x}^{(k-1)}} \mathcal{L} (\mathbf{x}^{(k)}|\mathbf{x}^{(k-1)})\textrm{Posterior}(\mathbf{x}^{(k-1)})$
       \EndFor
       \State $\textrm{Normalize}(\textrm{Prior}(\mathbf{x}^{(k)}))$
       \For{\textbf{all} $\mathbf{x}^{(k)}$}
       \State  $\textrm{Posterior}(\mathbf{x}^{(k)})\gets\mathcal{L}  (\mathbf{z}^{(k)}|\mathbf{x}^{(k)})\textrm{Prior}(\mathbf{x}^{(k)})$
       \EndFor
       \State $\textrm{Normalize}(\textrm{Posterior}(\mathbf{x}^{(k)}))$
        \State $\hat{\mathbf{x}}^{(k|k)}\gets \sum\limits_{\mathbf{x}^{(k)}}\mathbf{x}^{(k)}\textrm{Posterior}(\mathbf{x}^{(k)})$
        \State \Return $\hat{\mathbf{x}}^{(k|k)},\textrm{Posterior}(\mathbf{x}^{(k)})$
		\EndProcedure
    \end{algorithmic}    
	\caption{Bayesian estimator implementation.}\label{Bayesian}
\end{figure}

Intuitively, (\ref{prior}) is the prior belief in any point of target state space being current target state. It is our prior in the sense that it describes how likely we consider the target to be in any point in target state space, before considering the new measurements. Our prior belief in any point in target state space $\mathbf{x}^{(k)}$ depends on how likely it is for it to occur from all possible $\mathbf{x}^{(k-1)}$, and how strongly we believe in $\mathbf{x}^{(k-1)}$. Next, we update this conditional likelihood function using the new measurement:
\begin{multline}
\mathcal{L} (\mathbf{x}^{(k)}|\mathbf{z}^{(k)}, \ldots ,\mathbf{z}^{(0)})=\\ \mathcal{L}  (\mathbf{x}^{(k)}|\mathbf{z}^{(k-1)},\ldots,\mathbf{z}^{(0)}) \frac{\mathcal{L}  (\mathbf{z}^{(k)}|\mathbf{x}^{(k)})}{\mathcal{L}  (\mathbf{z}^{(k)}|\mathbf{z}^{(k-1)},\ldots ,\mathbf{z}^{(0)} )},
\label{posterior}
\end{multline}
where:
\begin{align}
\mathcal{L}  (\mathbf{z}^{(k)}&|\mathbf{z}^{(k-1)},\ldots ,\mathbf{z}^{(0)})=\nonumber\\
&\int \mathcal{L}  (\mathbf{z}^{(k)}|\mathbf{x}^{(k)})\mathcal{L}  (\mathbf{x}^{(k)}|\mathbf{z}^{(k-1)},\ldots,\mathbf{z}^{(0)}) \diff \mathbf x^{(k)}.
\label{finalBayes}
\end{align}
Intuitively, (\ref{posterior}) describes the posterior likelihood function, characterizing how new measurements affect our prior. If the new measurement $\mathbf{z}^{(k)}$ has high likelihood to observed, assuming the target state is $\mathbf{x}^{(k)}$, we increase the posterior belief $\mathbf{x}^{(k)}$ proportionally. If the measurements are unlikely to be observed at target state $\mathbf{x}^{(k)}$, we reduce our belief in $\mathbf{x}^{(k)}$ proportionally. For the next time step, we use $\mathcal{L} (\mathbf{x}^{(k)}|\mathbf{z}^{(k)}, \ldots ,\mathbf{z}^{(0)})$ to form our prior belief in $\mathbf{x}^{(k+1)}$.

The described recursive Bayesian estimation is mathematically optimal, yet we do not implement it as is due to multidimensional integrations in \eqref{eq:mmse}, \eqref{prior}, and \eqref{finalBayes}. We make slight adjustments in implementation by partitioning the target state space and by exclusively considering the region of interest. By partitioning the state space, we mean we quantize the values of states and tabulate the likelihoods in (\ref{eq:mmse}), (\ref{prior}), and (\ref{finalBayes}). By exclusively considering the region of interest, we mean we give zero prior to states that represent the target with more than max speed, and farther than the maximum distance. 

Fig.~\ref{Bayesian} represents our implementation of recursive Bayesian estimator. In this implementation, the integration over all state space in (\ref{prior}) is substituted with the summation over all quantized state space in third line of the Fig.~\ref{Bayesian}. This substitution is intended to reduce the computational expense of the algorithm by quantizing the state space. Similarly, the integration in (\ref{finalBayes}) is replaced with a one shot normalization in fifth line of Fig.~\ref{Bayesian}. The denominator in (\ref{posterior}) does not depend on the target state, and acts as a normalization factor to ensure that the posterior belief retains unit volume under the curve. Since we have quantized the state space, we can simply sum up the posterior belief over all the state space, and divide all the posterior beliefs by this constant to make sure the summation of the posterior beliefs remains equal to one. The expected value in (\ref{eq:mmse}) is also substituted with discrete expected value in tenth line of Fig.~\ref{Bayesian}. Overall by changing the number of quantized state space points (increasing resolution), we can trade off computation expense and estimation accuracy.

\begin{figure}
	\centering
	\includegraphics[width=0.45\textwidth]{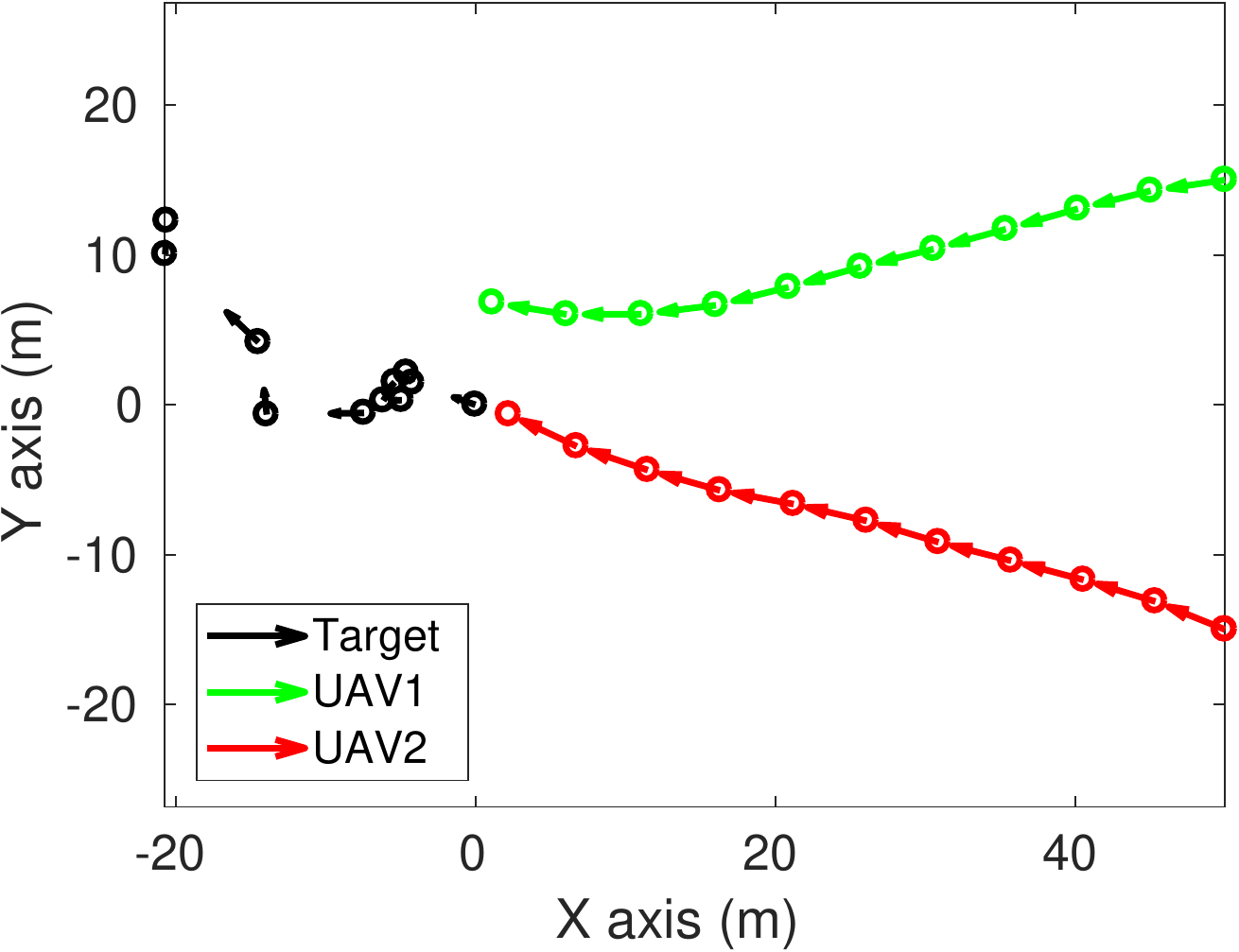}
	\caption{Trajectories of a target UAV and two tracking UAVs.}
	\label{1}
\end{figure}

\begin{table}
\caption{Default values of simulation parameters.}
\centering
\begin{tabular}{|c|c|c|}
\hline
Parameter & Description & Value\\
\hline
$N$ & Swarm size & 2\\
$v_\textrm{max}$& Maximum speed of tracking UAV  & 5 m/s\\
$q$& Probability of target not transmitting & 0.2\\
$\sigma_\textrm{th}$& Standard deviation of thermal noise &-80 dBm\\
$\sigma_\textrm{sh}$& Standard deviation of log-normal shadowing & 1\\
$Q$& Covariance of target movement process noise &diag(2)\\
$p_\text {on}$& Target transmission power & 30 dBm\\
$G$& Aggregated antenna gain & 1\\
 & Default path planning method &bio\\
 & Default estimation method &EKF\\
\hline
\end{tabular}
\label{default}
\end{table}
\begin{figure*}
       \centering
        \begin{subfigure}[b]{0.475\textwidth}
	\centering
	\includegraphics[width=\textwidth]{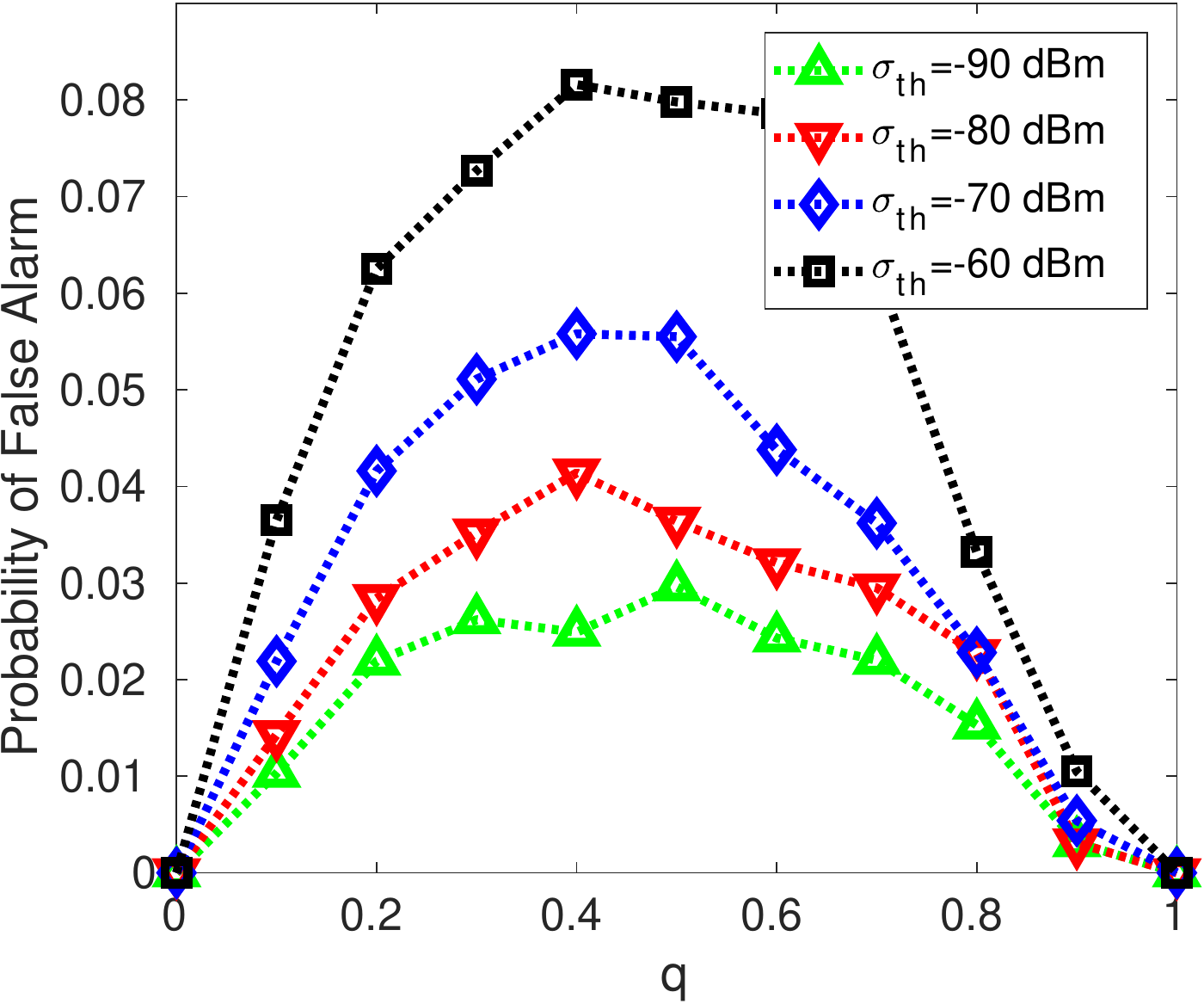}
	\caption{Probability of false alarm versus $q$, probability of target not transmitting.}
	\label{falseAlarm}
\end{subfigure}
\hfill
\begin{subfigure}[b]{0.475\textwidth}
	\centering
	\includegraphics[width=\textwidth]{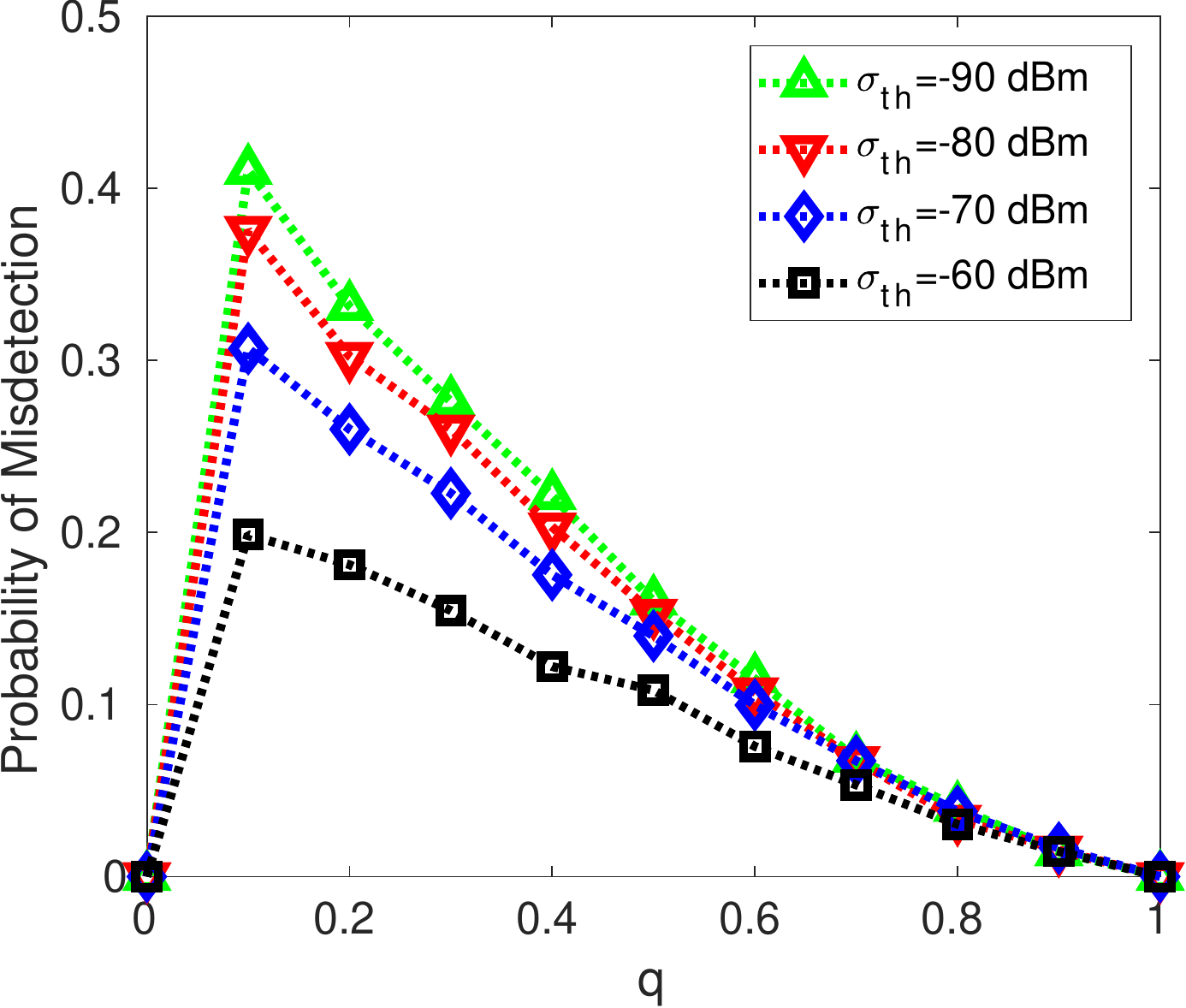}
	\caption{Probability of misdetection versus $q$, probability of target not transmitting.}
	\label{misDetection}
\end{subfigure}
\vskip\baselineskip
\begin{subfigure}[b]{0.475\textwidth}
	\centering
	\includegraphics[width=\textwidth]{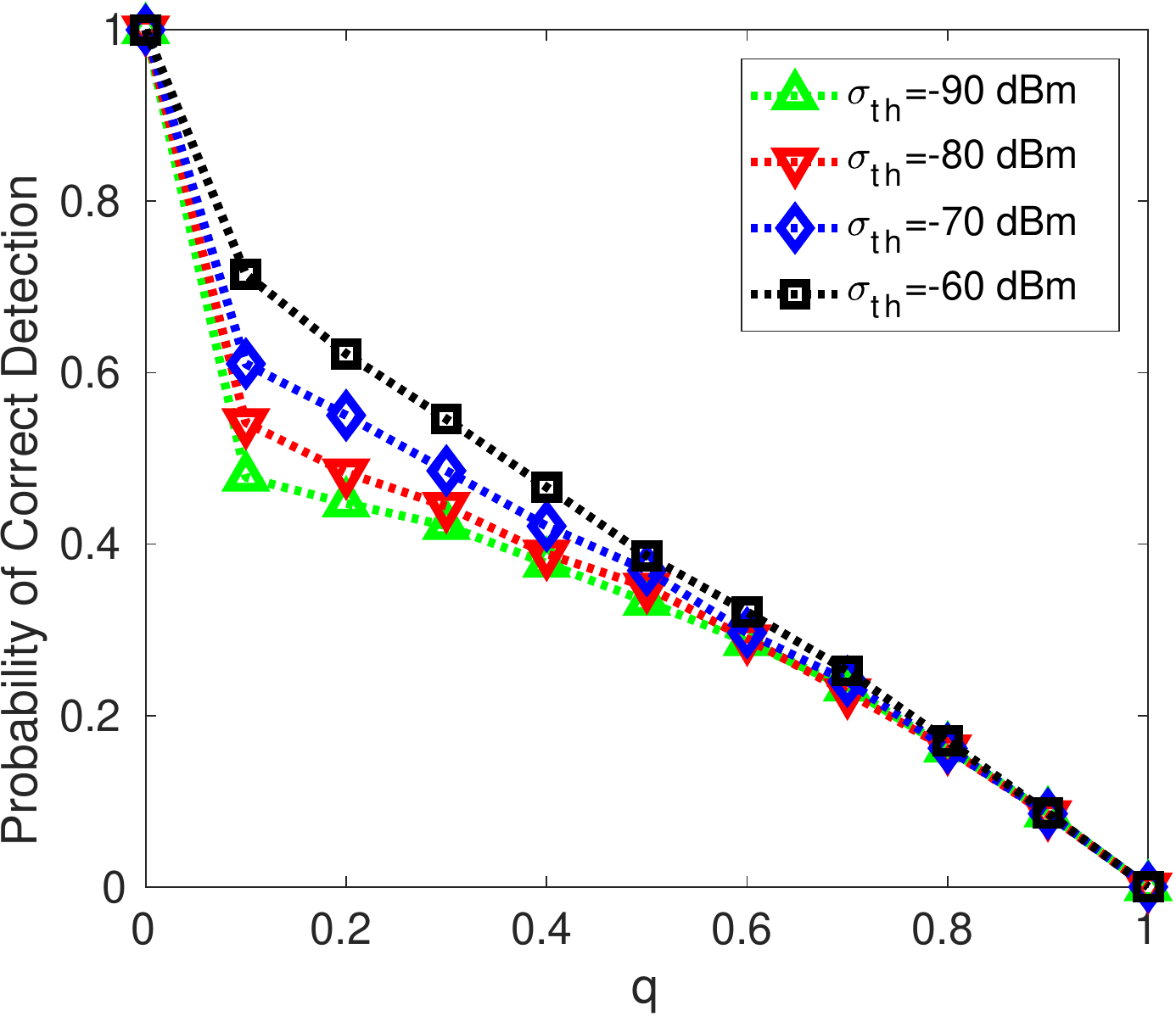}
	\caption{Probability of correct detection versus $q$, probability of target not transmitting.}
	\label{correctDetection}
\end{subfigure}
\hfill
\begin{subfigure}[b]{0.475\textwidth}
	\centering
	\includegraphics[width=\textwidth]{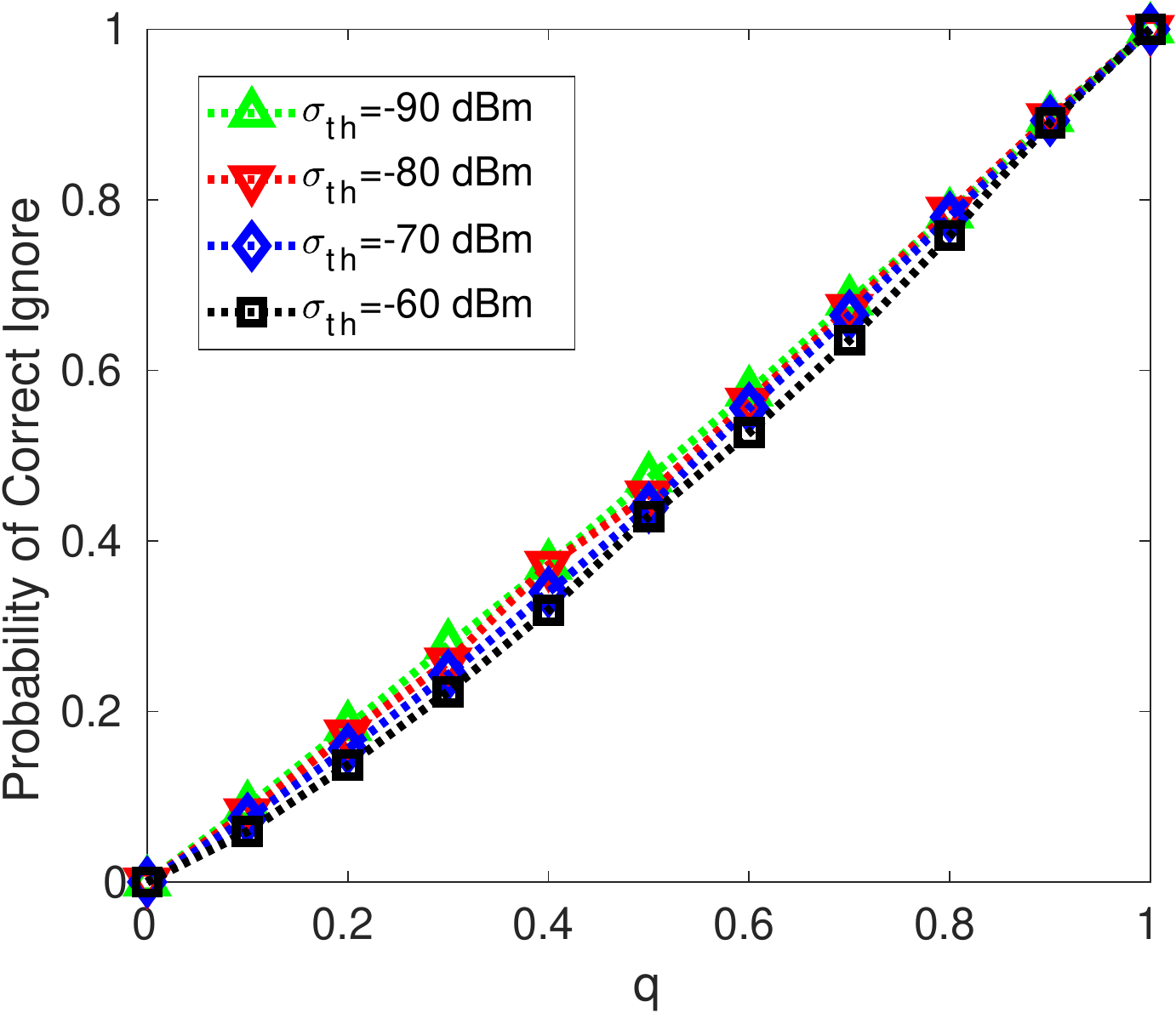}
	\caption{Probability of correctly ignoring the measurements versus $q$, probability of target not transmitting.}
	\label{correctIgnore}
\end{subfigure}
    \caption{Detection performance.}
    \label{detection}
\end{figure*}

\section{Simulation Results}
\label{sec:simulation}
In this section, we present our simulation results. In general at each scenario, the tracking UAVs use one of the path planning methods to navigate the area of interest, while using one of the estimation methods to update the target location that is used in their path planning method. Fig.~\ref{1} illustrates one instance of such simulation, with steepest descent method of path planning and detection based EKF estimation method. In Section~\ref{detectionError}, we investigate the performance of the optimal Bayesian cost detector in different noise scenarios. In Section~\ref{dCriterion}, we compare the resulting D-criterion of the two path planing approaches. In Section~\ref{estimatorError}, we investigate how changes in the stochastic parameters of the measurement process affect the error decay factor for EKF and Bayesian estimators. Finally in Section \ref{processingTime}, we present the computation time of each algorithm.

Unless otherwise noted, we use the default parameters that are presented in Table~\ref{default} to generate the simulation results. We assume target moves according to two dimensional double integral model. We also assume that the tracking UAVs are moving in the same two dimensional plane. It is worth mentioning that the methods presented in this paper are not restricted to two dimensions, and can be used in three dimensions without modification.
\begin{figure*}
	\centering
	\includegraphics[width=\textwidth]{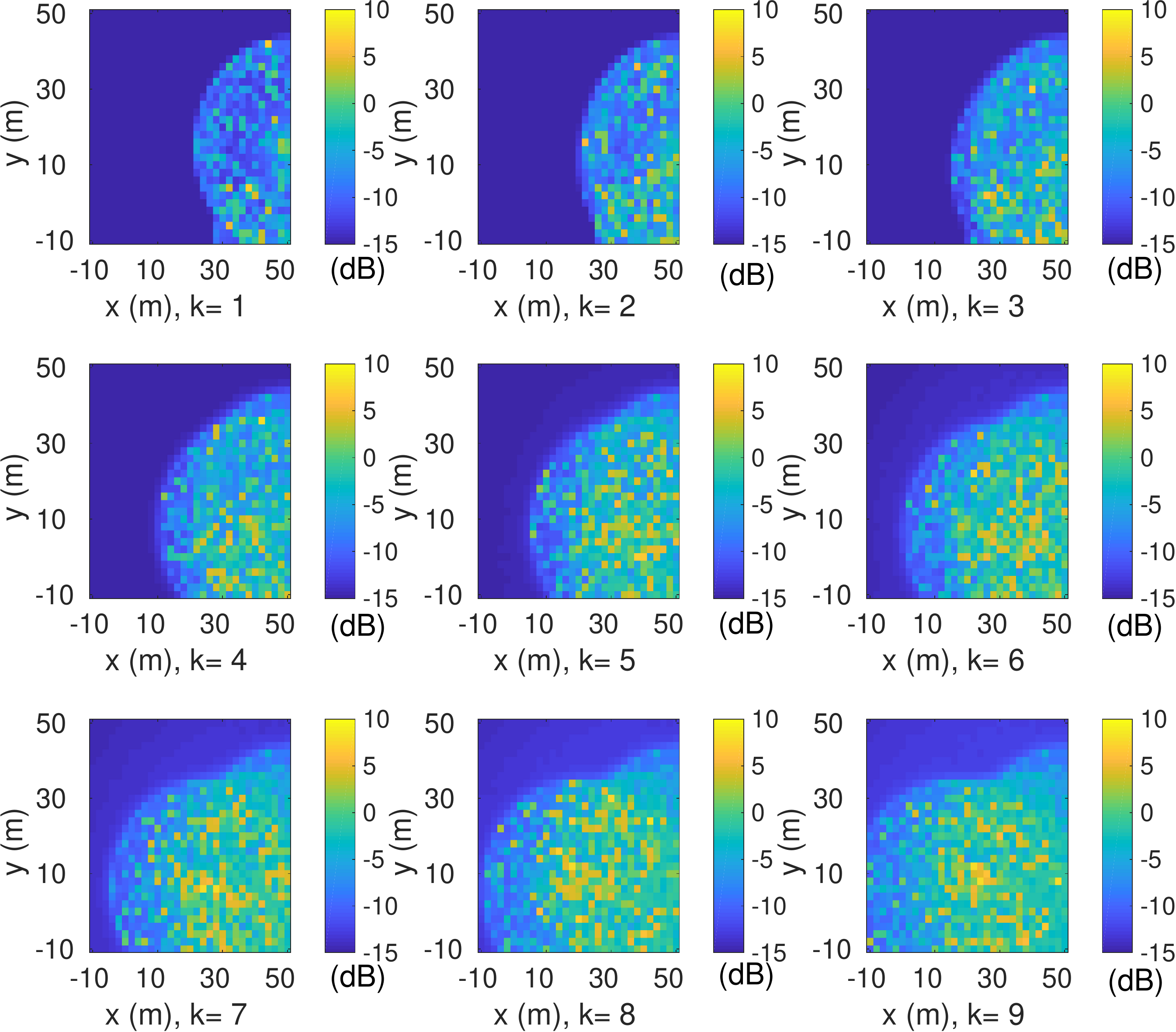}
	\caption{Evolution of the D-criterion (dB) over the entire region of interest at different time steps $k$.}
	\label{4}
\end{figure*}

\subsection{Detection Performance}
\label{detectionError}
We present, in Fig.~\ref{detection}, the performance metrics of the optimal Bayesian detector stage of the detection based EKF estimator, across wide range of thermal noise and probability of target transmission ($1-q$). These performance metrics are probability of false positive (Fig.~\ref{falseAlarm}), false negative (Fig.~\ref{misDetection}), correct detection (Fig.~\ref{correctDetection}), and correct ignore (Fig.~\ref{correctIgnore}). Obviously, when the probability of target transmission is zero or one, there is no error in detection and the detection becomes non-trivial when the transmission becomes probable but not certain. From Fig.~\ref{correctIgnore} we observe that all across the thermal noise range of -90 dBm to -60 dBm, the optimal detector has a very good performance in ignoring the measurements when there has not been a transmission from the target. On the other hand, from Fig.~\ref{correctDetection} we observe an interesting phenomenon in the detector performance in correctly detecting the transmission of the target when the target has a small chance of being silent. These two observations are consistent with false positive and false negative rates in Fig.~\ref{falseAlarm} (low false alarm across the range) and Fig.~\ref{correctDetection} (high miss when target has small chances of silence).

This becomes even more interesting when we notice that reducing thermal noise reduces false alarm, but increases the misdetection. Intuitively, and with a fixed threshold on likelihood ratio, we would not expect to see higher misdetection associated with lower measurement noise. Our investigation shows that when noise level is low, the optimal detector becomes conservative after acquiring a couple of measurements. That is to say, when the noise level is low and the target has low probability of silence, the prior information goes up rapidly, and the detector does not see any additional benefit in risking to consider barely confidence inspiring measurements. In this case, the optimal detector has so much prior information that it prefers to lose relatively small novel information content of the new measurements if the likelihood of the transmission of the target is not extremely high. In contrast, when noise is high or probability of target silence is high, the prior information is small and the detector is willing to consider any measurement that is likely to be generated from a transmitting target.

\subsection{D-Criterion}
\label{dCriterion}
The steepest descent path planning algorithm relies on estimation stage to localize the target. Subsequently, the steepest descent path planning maximized the D-criterion on that location by choosing the optimal path. The optimization does not consider the estimation error, and it may optimize the D-criterion on a location that is far from the location of the target, and as a consequence, the estimation stage will have an even harder job localizing the target. Naturally, we are interested to investigate how optimization of D-criterion over a single point trickles down the information to the vicinity of that point. Our simulations show that information in our particular setup is not very sensitive to localization error. For instance, Fig.~\ref{4} illustrates the available information over the entire region of interest as the UAVs are following steepest descent path planning. As the tracking UAVs navigate through the region of interest, they gain information about their environment. We note two observations: 1) the information gathered tends to fall rather rapidly over a certain radius; and 2) the information tends to be more or less homogeneous inside that radius.

\begin{figure}
	\centering
	\includegraphics[width=0.45\textwidth]{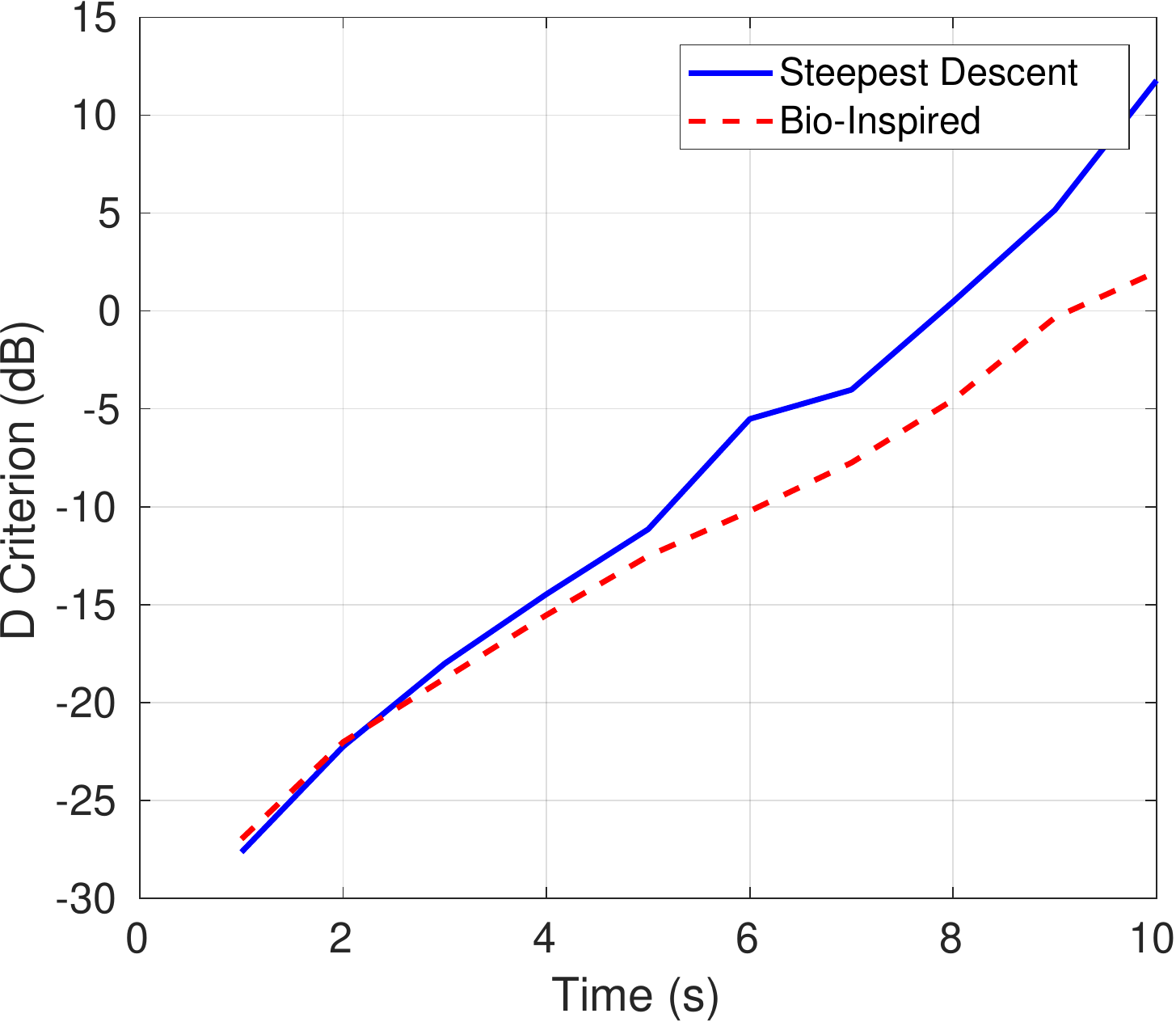}
	\caption{Time progression of the D-criterion for the steepest descent and bio-inspired path planning.}
	\label{2}
\end{figure}

Next we compare the time progression of D-criterion for bio-inspired path planning and steepest descent path planning. Fig.~\ref{2} depicts the result for $N=2$, $q=0.2$,  $\sigma^2_\textrm{th}=-60$ dBm, and $Q=\textrm{diag}(2)$. The D-criterion for both cases exhibit close to linear increase, suggesting an exponential increase in D-criterion. Although in the initial steps, the bio-inspired approach keeps up with the steepest descent approach, it starts to loose performance as the tracking UAVs get closer to the target. Such a behavior is expected, since as we get closer to the target, tracking UAVs can make better measurements if they start to encircle the target rather than following a straight path to the target. The performance hit is approximately 10 dB at the end of the simulation.

\begin{figure}
	\centering
	\includegraphics[width=0.48\textwidth]{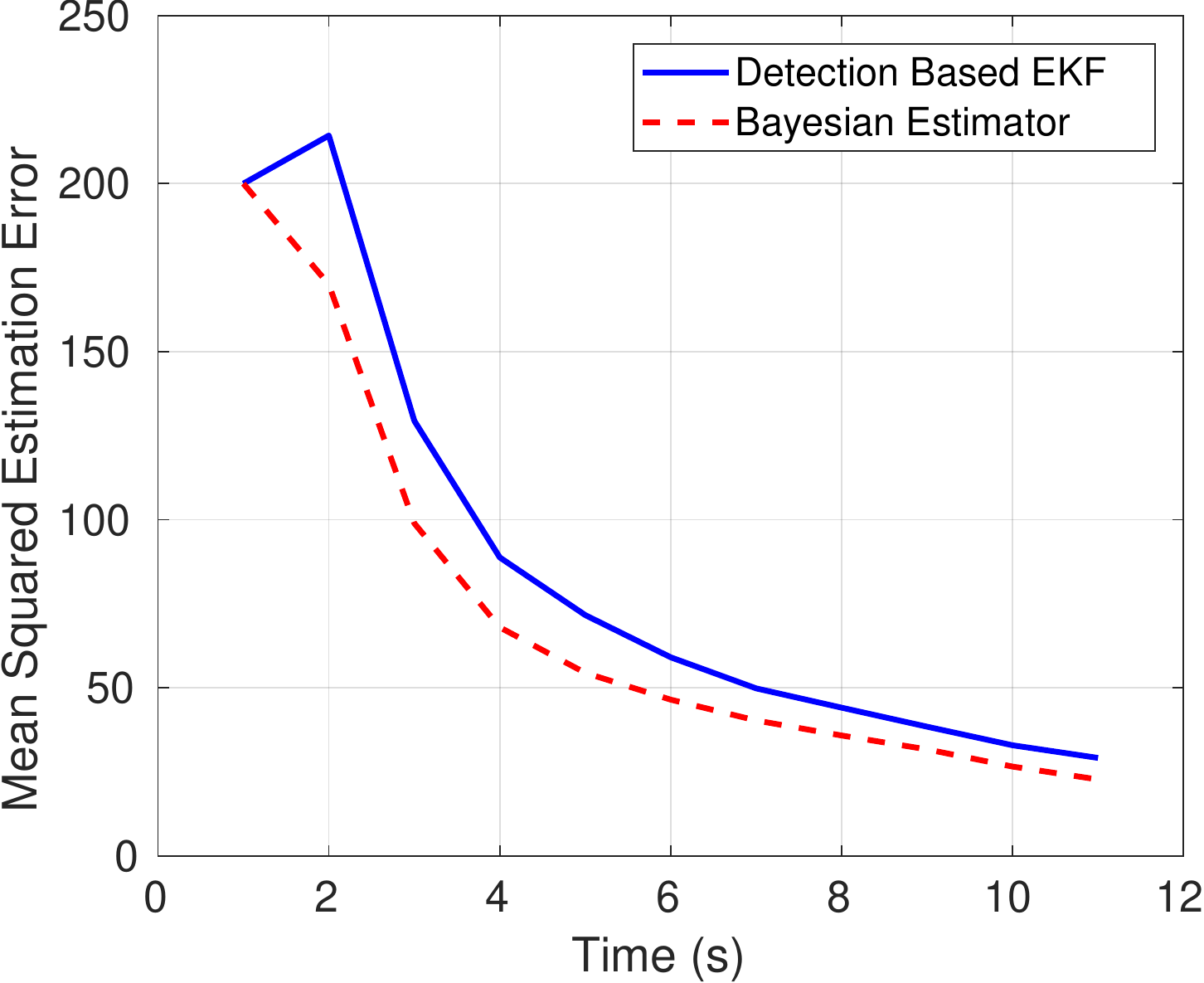}
	\caption{Time progression of the estimation error for the state vector ${\bf x}$ when using detection based EKF and Bayesian estimator.}
	\label{3}
\end{figure}

\subsection{Estimator Error}
\label{estimatorError}
In Fig.~\ref{3}, we compare the mean square estimation error of detection based EKF and Bayesian estimator.  We show the result for $N=2$, $q=0.2$,  $\sigma^2_\textrm{th}=-60$ dBm, and $Q=\textrm{diag}(2)$. The mean squared estimation error converges below 50 for both estimators, with detection based EKF showing an increase in the first update. This increase increase is due to second order difference equations that govern the process model. In this case, the errors in initial speed estimation will carry into the next step. We note that by mean squared estimation error, we mean

\subsection{Processing Time}
\label{processingTime}
Although our implementation of the aforementioned algorithms are by no means computationally optimal, we provide the run times that we have observed to give a rough idea about the computational complexity of each approach. All codes are implemented using MATLAB, and are executed on Intel Core~i7 6700k. The bio-inspired path planning and detection based EKF run time is on the order of measurement accuracy of MATLAB (1 ms) for the entire simulation. On the other hand, Bayesian estimation processing time takes 30.8 seconds of runtime for each simulation. Bayesian estimation processing time standard deviation is observed to be 5.6 seconds. This result is obtained by subdividing the target state space into 10000 regions. The processing time can be significantly improved by considering a lower order state space for the target, and/or lower number of quantized subspaces. Mean steepest descent path planning processing time is measured to be 41.3 s for the entire simulation. Steepest descent path planning processing time standard deviation is measured to be 9.8 seconds. We have used exhaustive optimization to find the steepest path since we wanted to trade off extra computation for better D-criterion. The performance may be significantly improved if other types of non-convex optimization are used.  

\section{Conclusion}
\label{sec:conclusion}
We have decomposed the challenge of tracking a moving intermittent RF source into two distinct phases of path planning and estimation. We have derived steepest descent path planning to address the first challenge, and developed detection based EKF and recursive Bayesian estimator to address the second challenge. We have compared the performance of steepest descent path planning with a heuristic bio-inspired path planning and have shown that steepest descent method continues to gather more information compared to bio-inspired path planning method. The available information through steepest descent path planning is 10 times more than the bio-inspired path planning after only 10 seconds of data gathering. We have shown that this increase in performance comes at the price of higher computational cost. We have also compared the performance of detection based EKF with recursive Bayesian estimator. We have shown that Bayesian estimator marginally outperforms detection based EKF estimator in terms of localization error, but at a larger computational cost. 

\section*{Acknowledgement}

This research was supported in part by the National Science Foundation under
the grant number CNS-1446570 and by the National Aeronautics and Space Administration under the grant number NNX17AJ94A.

\bibliographystyle{IEEEtran}
\bibliography{IEEEabrv,main}
\end{document}